%% file: main.tex
\documentclass[a4paper,UKenglish,cleveref, autoref, thm-restate]{lipics-v2021}

\hideLIPIcs  

\nolinenumbers

\usepackage[graphicx]{realboxes}
\usepackage{fontawesome5}
\usepackage{placeins}
\usepackage{pifont}
\usepackage{rotating}
\usepackage{tikz,collcell}
\usepackage{balance}
\usepackage{lipsum}
\usepackage{rotating}
\usepackage[export]{adjustbox}
\newcommand*{\cellseta}{\pgfqkeys{/cell}}
\newcommand*{\myCell}{\cellseta}
\newcolumntype{C}[1]{>{\collectcell\myCell}#1<{\endcollectcell}}
\makeatletter
\tikzset{overlay linewidth/.code=\tikz@addmode{\tikzset{overlay}}}
\cellseta{.unknown/.code={\edef\pgfkeys@temp{\noexpand\cellset{box=\pgfkeyscurrentname}}\pgfkeys@temp}}
\makeatother
\cellseta{
  head/.code=\rlap{\scriptsize\rotatebox{90}{#1}}\hspace*{1.2em},
  myCell/.style={
    draw=black,
    overlay linewidth,
    inner sep=+0pt,
    outer sep=+0pt,
    anchor=center,
    fill={#1},
    minimum size=+1.2em},
  box/.code={%
    \tikz[baseline=-0.5ex]
      \node[/cell/myCell={#1}]{};},
  box/.default=none,
  ./.style={box},
  @define/.style args={#1:#2}{#1/.style={box=#2}},
  @define/.list={%
    technical:cyan,
    chi:   green, bribe:  red,
    buy:          orange, todo:       yellow}
  }
\usepackage[most]{tcolorbox}
\tcbuselibrary{skins,breakable}

\usepackage{pifont}
\usepackage{makecell}
\usepackage{lscape}
\usepackage{threeparttable}
\usepackage{amsmath}
\usepackage{amssymb}
\usepackage{afterpage}
\usepackage{xurl} 
\usepackage{tabularray}
\usepackage{lmodern,babel,adjustbox,booktabs,multirow}

\usepackage{caption}
\usepackage{subcaption}

\theoremstyle{definition}
\newtheorem*{ndef}{Definition}

 \newtcolorbox{titlebox}[5]{enhanced,center,colframe=black,colback=white,boxrule={#3},arc={#2},auto outer arc,%
 breakable,pad at break*=5pt,vfill before first,before={
 },before={\par\smallskip},after={\par\smallskip},top=12pt,left=4pt,%
 enlarge top by=2pt,
 fontupper=\small,
 title={\rule[-.3\baselineskip]{0pt}{\baselineskip}\normalsize\sffamily\bfseries #1}, varwidth boxed title*=-30pt, 
 attach boxed title to top left={yshift=-10pt,xshift=10pt}, coltitle=black,
 boxed title style={colback=white,boxrule={#5},arc={#4},auto outer arc}
 }

 \newenvironment{casestudybox}[1]
 {\begin{titlebox}{Case Study \normalfont #1}{0.5pt}{0.5pt}{1pt}{0.75pt}}
 {\end{titlebox}}
\NewTableCommand\SCC[1]{\SetCell{bg=#1}}
\newcommand\T[1]{\vspace{4pt}\noindent\textbf{#1} }

\title{SoK: Attacks on DAOs} 

\author{Rainer Feichtinger\footnote{The authors of this work are listed alphabetically, correspondence through \url{daoattacks@ethz.ch}.}}{ETH Zurich}{rainerfe@student.ethz.ch}{https://orcid.org/0000-0001-5075-0959}{}
\author{Robin Fritsch}{ETH Zurich}{rfritsch@ethz.ch}{https://orcid.org/0009-0006-3123-1735}{}
\author{Lioba Heimbach}{ETH Zurich}{hlioba@ethz.ch}{https://orcid.org/0000-0002-8258-1712}{}
\author{Yann Vonlanthen}{ETH Zurich}{yvonlanthen@ethz.ch}{https://orcid.org/0000-0001-5736-8197}{}
\author{Roger Wattenhofer}{ETH Zurich}{wattenhofer@ethz.ch}{https://orcid.org/0000-0002-6339-3134}{}

\authorrunning{R. Feichtinger, R. Fritsch, L. Heimbach, Y. Vonlanthen, R. Wattenhofer} 

\Copyright{CC-BY} 



\category{} 

\supplementdetails[subcategory={Website}, cite={}, swhid={}]{Dataset}{https://daoattacks.ethz.ch}

\acknowledgements{We thank Hubert Ritzdorf from ChainSecurity for his precious feedback.}

\EventEditors{Rainer B\"{o}hme and Lucianna Kiffer}
\EventNoEds{2}
\EventLongTitle{6th Conference on Advances in Financial Technologies (AFT 2024)}
\EventShortTitle{AFT 2024}
\EventAcronym{AFT}
\EventYear{2024}
\EventDate{September 23--25, 2024}
\EventLocation{Vienna, Austria}
\EventLogo{}
\SeriesVolume{316}
\ArticleNo{16}

\begin{document}

\maketitle

\begin{abstract}
\textit{Decentralized Autonomous Organizations (DAOs)} are blockchain-based organizations that facilitate decentralized governance. Today, DAOs not only hold billions of dollars in their treasury but also govern many of the most popular \textit{Decentralized Finance (DeFi)} protocols. This paper systematically analyses security threats to DAOs, focusing on the types of attacks they face. We study attacks on DAOs that took place in the past, attacks that have been theorized to be possible, and potential attacks that were uncovered and prevented in audits. For each of these (potential) attacks, we describe and categorize the attack vectors utilized into four categories. This reveals that while many attacks on DAOs take advantage of the less tangible and more complex human nature involved in governance, audits tend to focus on code and protocol vulnerabilities. Thus, additionally, the paper examines empirical data on DAO vulnerabilities, outlines risk factors contributing to these attacks, and suggests mitigation strategies to safeguard against such vulnerabilities. 
\end{abstract}

\section{Introduction}
\textit{Decentralized Autonomous Organizations (DAO)} are organizational structures that facilitate the trustless management of projects that run on a blockchain~\cite{bell2021blockchain}. In DAOs, governance is typically controlled by the holders of a designated governance token. Those who own these tokens can thus determine the course of the DAO. Today, DAOs govern various blockchain projects, such as ecosystem governance of Layer 2s (e.g., Arbitrum and Optimism) and many of the most-used decentralized applications (e.g., Aave, Compound, ENS, Lido, MakerDAO, and Uniswap). Moreover, DAOs are estimated to hold and control in excess of \$30B in their treasuries~\cite{2023deepdao}. Consequently, they hold significant power and a central position in the blockchain ecosystem.

DAOs have been threatened by attacks and hacks ever since their inception.
\textit{``The DAO''} on Ethereum was the first attempt at creating a DAO on a blockchain. However, in 2016 an infamous hack stole \$50M worth of ETH from the DAO before it even became operational~\cite{2023thedao}. The event was so severe that it led to a controversial hard fork of the Ethereum blockchain. The original (unforked) blockchain still operates today and is known as Ethereum Classic. Notably, even before the fatal hack, other possible attacks on The DAO had been discussed \cite{mark2016thedao}. The DAO hack highlights the significant threat attacks on DAOs can present not only to the DAOs themselves but also to the broader ecosystem. Additionally, given that DAOs are still in their early days and the ongoing evolution of their design frameworks, DAOs are particularly vulnerable to various novel attack vectors. 

In this work, we study real-world incidents and attacks on DAOs, attacks that have been theorized to be possible, and new potential attack vectors. We summarize our main contributions in the following. 
\begin{itemize}
    \item We present and categorize attack vectors on DAOs. To be precise, we categorize attacks on DAOs into four categories: (i) bribing (BR) attacks, (ii) token control (TC) attacks, (iii)  human-computer interaction (HCI) attacks, and (iv) code and protocol vulnerability (CP) attacks.
    \item We examine 28 real-world incidents and attacks across four blockchains and indicate the attack vectors utilized. Our work finds that these attacks exploited vectors from all four introduced categories fairly evenly. Similarly, we categorize attacks described in academic papers or reports as well as those uncovered and prevented through audits. Notably, less tangible attack vectors that take advantage of human and economic aspects involved in governance represent a majority of real-world incidents but are generally not analyzed in audits, which heavily skew toward code and protocol vulnerability attacks. 
    \item Guided by our categorization of real-world incidents, we introduce seven risk factors for DAOs and empirically analyze the susceptibility of 26 DAOs of all shapes and sizes to them.
    \item Finally, we collect and discuss various mitigations and safeguards for DAOs.
\end{itemize}

With our work, we aim to enhance the understanding of DAO security challenges and guide the development of more robust governance frameworks.

\section{Background}

An early definition of DAOs was provided by Vitalik Buterin, who argued that DAOs are entities with internal capital (i.e., \emph{treasuries}), that have automated processing at their core, and human processing at their edges~\cite{buterin2014daodefinition}.

This definition still holds true today, though DAOs have been continually evolving and are still undergoing substantial efforts to improve. Therefore, countless implementations exist. Nonetheless, some specific designs have reached greater popularity, as new DAOs borrow ideas, pieces of code from the smart contracts implementing the DAOs logic, or entire structures from pre-existing DAOs.  Examples thereof are the OpenZeppelin implementations based on the Compound Governor contracts (also used e.g., by Uniswap, ENS, and Gitcoin) and Aragon DAOs (used by e.g., Lido and Curve). Other DAOs might follow a more unique design (e.g., Maker and Optimism) or reuse code only partially. In the following, we aim to distill the main attributes that current DAOs share.

\textbf{Token Voting.} Reaching an agreement between individuals in the decentralized setting of a blockchain is not as simple as in the physical world, where identities are known. On blockchains, only pseudonymous addresses are publicly available. In particular, these addresses could include \textit{sybils}, i.e., multiple addresses created and controlled by a single individual.
To address this issue, DAOs typically issue governance tokens. These tokens come with voting rights.\footnote{Generally, each token counts as one vote. Curve is an exception, where the voting power depends not only on the number of tokens held but also the duration the tokens are locked for.} Initially, these tokens can be distributed among stakeholders through various methods. The most popular methods include giveaways (\textit{airdrop}) to early users of a protocol, as well as allocating a portion of the tokens to the development team or early investors. After launching, the governance tokens become freely tradable on the open market, enabling individuals to acquire them and thus acquire voting power. In this regard, governance tokens exhibit parallels with shares in companies that grant holders voting rights at shareholder meetings.

\textbf{Voting Rights and Delegation.} While token holders directly vote on each proposal themselves in some DAOs (e.g., Lido), other DAOs introduce an intermediary in the voting process: delegates. In these DAOs, token holders can delegate the voting power associated with their tokens to a delegate they believe represents their views well. Inspiration for this \emph{delegative} approach is taken from \emph{liquid} democracy \cite{ford2002delegative, behrens2017liquid}. Proponents of this approach for DAO voting argue that delegation allows token holders to participate in the governance process passively (i.e., they must not actively keep track of each proposal) and can further help reduce the blockchain transaction fees incurred by DAO members. While delegation is optional in some DAOs (e.g., Aave), many DAOs require delegation (e.g., Compound, ENS, and Uniswap). In DAOs requiring delegation, before tokens can be used for voting, the address holding the tokens must delegate them to a \emph{delegate} address (which may be the same address). Finally, some DAOs require the tokens to be locked in a specified contract to gain the associated voting rights (e.g., Curve and Maker).  

\textbf{Deliberation Phase.} Generally, DAO governance processes involve multiple steps, which could include discussions on public governance forums or off-chain \textit{temperature check} votes before a final vote~\cite{decrypt2021whatissnaphot} ahead of a proposal being put forward on-chain and voted on. However, these steps are often mere social conventions. The final on-chain vote is often the only obligatory part of the process.

\textbf{Proposals.} Generally, any token holder with a sufficient number of tokens (exceeding a pre-defined proposal threshold) can put forward a proposal. There are some DAOs though that force proposals to be vetted by a committee before being accepted on-chain. 

\textbf{Voting Phase.} Once a proposal is submitted on-chain, there is generally a \textit{proposal delay}, i.e., a pre-specified number of blocks before a snapshot is taken of the balances of all token holders (or delegates). The snapshot determines their voting power, and cannot be changed a posteriori. Thereafter, voting starts and generally lasts for a pre-determined number of blocks. During voting, any address with voting rights (token holder or delegate) may submit a vote. 

\textbf{Execution Phase.} If a majority votes in favor of a proposal and a pre-determined Quorum is reached, the proposal is accepted. In some DAOs, on-chain proposal execution is automatic, potentially only after a pre-defined \textit{timelock delay}. The execution is scheduled manually by a trusted party in other DAOs.

\section{Categorization of Attack Vectors}\label{sec:attack_vectors}

In the following, we provide a categorization and description of attack vectors. 

\subsection{Bribing}\label{sec:bribe}

\begin{ndef}
In a \emph{bribing (BR)} attack, an attacker pays to change votes or to acquire voting power without acquiring the underlying governance tokens. The controlled votes and voting power are then utilized to pass a malicious proposal in a governance vote.
\end{ndef}

Bribing attacks can take the form of paying to obtain voting rights of governance tokens to vote for a certain proposal without acquiring the underlying token, which is often referred to as \textit{vote buying}. Another possibility is directly bribing token holders or delegates.

Given that vote buying has been documented in shareholder governance of traditional companies~\cite{lan2005sale}, it is a plausible future concern for DAO governance and was already been a topic of discussion since the early days of DAOs~\cite{daian2018dark, buterin2021voting}.

\T{Bribing Token Holders or Delegates (BR1).}
Bribing governance participants, i.e. token holders or delegates, can take many forms: it can be done on-chain or off-chain, programmatically using smart contracts or by personal contact. Furthermore, bribes could be fixed sums or a proportion of the proceeds from a successful attack. Note that using proceeds of the attack to bribe leads to a situation similar to an attack by a majority coalition, where the proceeds are split among participants (see TC5).

When voting power is highly centralized, as is the case for many DAOs at the time of writing~\cite{feichtinger2023shortcomings}, bribing only a few of them can suffice to change a vote. On the other hand, voting rights being highly distributed can also make it cheaper for an attacker to bribe: holders of small amounts of voting power, besides having little to lose from a successful attack, also have little influence on the outcome of a vote. Hence, it can be economically rational for them to cheaply sell their vote, as described by Buterin~\cite{buterin2017voting}.

Bribing delegates to vote a certain way could potentially be particularly attractive for an attacker. For governance systems using delegated token voting, a small number of delegates often controls large amounts of voting rights. On the other hand, these delegates do not actually hold the corresponding amount of governance tokens, meaning they are not exposed to the price risk from a successful governance attack. Hence, bribing them could potentially be significantly cheaper for an attacker than bribing governance token holders.

One deterrent against a delegate bribing attack, that is present in most current DAOs, is the fact that most delegates are often publicly (or at least pseudonymously) known. This means that delegates stand to lose their reputation and future earnings based on it, and may even face a risk of criminal charges for accepting bribes.

\T{Vote Buying Protocols (BR2).}
The act of vote buying or \textit{bribing} can be facilitated by a smart contract protocol. Such protocols allow token holders to deposit their governance tokens into pools, and earn fees from users paying to use the voting rights of the pooled tokens. In particular, this means that vote buyers do not need to deposit collateral, contrary to using traditional lending platforms such as Compound (see Section~\ref{sec:acquisition}). Paladin Lending, as one example of a vote buying protocol, is described in the following.

\begin{casestudybox}{Paladin Lending}
    Paladin Lending~\cite{2023paladin} lets holders deposit their tokens into pools and in return receive a proportional share of the fees collected in the pool. Users can then borrow the voting power of deposited tokens.
    A loan contract is automatically created if a user wants to borrow voting power. The borrowed token amount is transferred from the pool to the loan contract, and the votes are delegated to the user. Hence, the user has no direct access to the tokens but can use their voting power. The user pays a fee for borrowing the voting power. At the latest when this fee has been consumed, the tokens in the loan contract will be returned to the pool.
\end{casestudybox}

Importantly, with a vote buying protocol such as Paladin Lending, one does not borrow the actual token, but only the voting rights. We also perform an empirical analysis of Paladin Lending (see 
Appendix~\ref{app:paladin})
to show that low liquidity currently does not allow attacks exclusively using this attack vector.

Daian et al.~\cite{daian2018dark} introduce a particular type of vote buying protocol: \emph{Dark DAOs}. In addition to facilitating vote buying using smart contracts, Dark DAOs are implemented in a privacy-preserving manner. Note that activity on vote buying protocols such as Paladin Lending is publicly recorded on the blockchain. Vote buying activities through Dark DAOs, on the other hand, cannot be detected, meaning that other governance participants cannot react to such an attack. While there are no known cases of active Dark DAOs at the time of writing, they have been theoretically studied by Austgen et al.~\cite{austgen2023dark}, and proof-of-concept prototypes have been published~\cite{augsten2024dark}.

\subsection{Token Control}\label{sec:acquisition} 

\begin{ndef}
    With \emph{token control (TC)} attacks, an attacker takes possession or is already in possession of a significant amount of governance tokens. The attacker then uses the voting power associated with these tokens to get their malicious proposal accepted in a governance vote.
\end{ndef}
This family of attack vectors is of the simplest nature. The attacker merely gains control of a sufficient number of governance tokens to take over the DAO by passing a malicious proposal according to DAO's intended voting process. 

Depending on the governance model implemented by the DAO, the required proportion of governance tokens for a successful attack varies. For instance, many DAOs require tokens to be delegated to an address for them to be used in voting and take a snapshot of the current state of delegations at the start of the voting period.
For such governance systems, an attacker must only hold a token amount exceeding the amount of previously delegated tokens, and delegate these governance tokens to themselves, thereby securing a majority of the delegated votes. By timing the creation of a proposal accordingly, an attacker can leave very little time (depending on the governance system's parameter choices, see RF4 in Section~\ref{sec:riskfactors} for more details) for others to react and delegate their tokens. This can almost guarantee the attacker the required voting power to pass their desired proposals. For DAOs that do not require the tokens to be delegated, the attacker would need to hold more than 50\% of the circulating token supply for a guaranteed victory of their proposal or hope that not sufficiently many votes are cast, i.e., voter turnout does not increase dramatically in face of a malicious proposal. Short voting windows as well as the absence of reliable communication channels further increase the risk of such attacks for these DAOs.

In the following, we discuss the main possibilities for an attacker to gain possession of the required voting power. Note that in Section~\ref{sec:riskfactors}, we provide an additional empirical analysis of the susceptibility of a set of 26 DAOs to this kind of attack.

\T{Token Purchase (TC1).} The attacker buys governance tokens on the open market. This can be done on-chain through decentralized exchanges, or on off-chain centralized exchanges. After using the tokens for voting, the attacker can sell back the tokens to the open market. Importantly, when buying governance tokens, the attacker takes on price risk while holding the tokens. If the attack leads to a decrease in the governance token's market price, the attacker incurs a financial loss. Additionally, the attacker pays trading fees when buying and selling the tokens. Note that the attacker can potentially hedge the price risk using derivatives. However, the availability of such derivatives may be limited depending on the governance token in question.

Attacks through token acquisition have been attempted and have occurred in several DAOs. They are especially attractive and profitable if the value of the treasury (excluding the governance token itself, which is likely to decrease in value in the event of an attack) exceeds the capital required to buy the necessary voting power. In Section~\ref{sec:riskfactors}, we compare the treasury values of DAOs to the value of delegated tokens for a set of 26 DAOs. It is relatively common for the total value of the DAO's treasury to exceed the value of delegated tokens, though this is only rarely the case when excluding the governance tokens from the treasury.

In the following, we present a case study of two consecutive recent governance attacks through token acquisition on the Indexed Finance DAO, a protocol for portfolio management. While interest in the project declined after it was hacked in October 2021~\cite{2023indexed}, various tokens remained in the project's timelock contract controlled by the DAO. 

\begin{casestudybox}{Indexed Finance} 
    On 16 November 2023, over ten hours, the attacker bought NDX tokens (i.e., the protocol's governance token) via decentralized exchanges, self-delegated these tokens, initiated a proposal, voted in favor of this proposal, and sold the tokens again~\cite{2023indexed}. The proposal would allow the attacker to take control of the timelock, mint new NDX tokens, and steal tokens from the timelock (including both NDX and other tokens). A call for action by one of the protocol founders asked users to vote against the proposal. In the end, user votes against the proposal were sufficient to narrowly prevent the attack. Interestingly, the attacker sold his NDX tokens before the end of the proposal and thereby lost his voting power. As a result, the proposer would have been below the proposal threshold and the proposal could have been canceled by anyone. However, this was not done.
    
    Fearing a potential second attack, the community attempted to implement defensive measures. They created a proposal to transfer control of the timelock to a smart contract not be under anyone's control, i.e., the tokens in the timelock would forever be inaccessible if the proposal were executed. Then, on 22 November 2023, another attacker (i.e., a different account than the previous attacker) created a similar proposal that would transfer the admin rights of the timelock to the attacker. This time, the attacker acquired more NDX tokens than the 16 November attacker, and there were not enough votes against this proposal. Thus, the only way to stop the attacker from getting access to the tokens was through passing the proposal that would make the tokens forever inaccessible. Importantly, as this proposal was created a day earlier, it would not only execute first but the attacker also only acquired the tokens after voting had started on the community's proposal and therefore did not have the majority in that vote. What followed, as no one wanted the community's proposal to be executed, was a message exchange between the attacker and the Indexed Finance team using input data of Ethereum transactions. In the end, an agreement was reached, and the attacker received $\approx$ \$10K via an escrow contract after withdrawing his proposal. In conclusion, the two attacks were only mitigated by luck (i.e., the first attacker bought too few tokens) and by unorthodox proposals (i.e., making the tokens forever inaccessible).

    In the aftermath of the attacks, the Indexed Finance DAO accepted a proposal that transferred control of the timelock to a multi-signature wallet controlled by former protocol contributors. 
\end{casestudybox}

Indexed Finance demonstrates the complexities of protecting against this attack vector in the absence of adequate countermeasures. Nevertheless, there exist potential protections that DAOs can put in place. For example, DAOs may opt to restrict proposals from spending the entire treasury or grant veto power to a multi-signature. We provide more detail in Section~\ref{sec:safe}.  

\T{Token Loan (TC2).} The attacker borrows governance tokens against collateral using lending protocols. Apart from needing to post collateral, the attacker also pays borrowing fees for the period of borrowing the tokens. Importantly, the attacker does not take on price risk when borrowing tokens. After voting for an attacking proposal, the full amount of governance tokens can be returned and the attacker receives back their collateral. 

There have been several alleged attempts of DAO attacks through token loans. In early 2022, Justin Sun presumably borrowed large amounts of MKR, the governance token of MakerDAO, to sway a vote. However, he returned the tokens after his actions were detected and did not end up voting~\cite{2022makerdaoc}. A couple of days later, a similar failed attempt by Justin Sun took place in Compound's governance with borrowed COMP tokens~\cite{2022compound}.

\T{Flash Loan (TC3).} With a flash loan, the attacker only borrows the governance tokens for the duration of a transaction. While the attacker pays a fee to borrow the governance token, the attacker does not need to post any collateral, i.e., does not require access to significant funds. Many protocols protect themselves against flash loan attacks by implementing a delay between the proposal creation and the start of the voting period. Nonetheless, flash loan attacks on DAOs have occurred in the past, the most prominent example is a flash loan attack on the Beanstalk governance described in the following case study. 

\begin{casestudybox}{Beanstalk}
	Beanstalk is a stablecoin protocol. On 17 April 2022, Beanstalk suffered an attack that resulted in damages of approximately \$182M, netting the attacker a profit of around \$76M~\cite{dotan2023vulnerable,faife_how_2022,beanstalk2022}. The attacker exploited a vulnerability in Beanstalk's governance system, which was not secure against flash loan attacks. The attacker took a flash loan worth approximately \$1B. This loan allowed them to achieve a two-thirds majority in Beanstalk's governance. With this majority, they could execute a malicious proposal immediately using an emergency commit function.
\end{casestudybox}\vspace{-6pt}

\T{Whale Activation (TC4).} Inactive token holders with a large number of tokens (often referred to as whales) can suddenly become active in the governance. In DAOs requiring tokens to be delegated, this can be especially problematic. An attacking whale can delegate their tokens and promptly initiate a proposal. Importantly, large entities holding sufficiently many tokens to take over the DAO exist for many DAOs using delegated token voting (see Section~\ref{sec:riskfactors}). Notably, there was one instance in the past where a centralized exchange unexpectedly delegated the UNI governance tokens it held, i.e., the tokens custodied on behalf of its users. They, however, claimed to have accidentally delegated these tokens~\cite{2022binance}.

\T{Majority Coalition (TC5).}
In governance systems using majority token voting, it is generally possible for a simple majority of voting tokens to accept any proposal, and effectively, take control of the DAO. In particular, the majority could distribute the entire DAO treasury among themselves. Settings of this type have been modeled in game theory as \textit{coalition games with transferable utility} or \textit{majority games} with \textit{stable sets} describing possible attacking coalitions~\cite{buchanan1961majority, jordan2003majorityrule}. Such coalition attacks are specifically attractive when the treasury value of a DAO is high compared to the value of (delegated) governance tokens. We have empirically studied this relation in Section~\ref{sec:riskfactors} (see RF3) for 26 DAOs.

Of course, a majority of voting tokens can also vote to split the treasury among \emph{all} token holders, or more generally, dissolve the DAO. In this particular case, i.e., if all token holders get a share of the treasury proportional to their voting power, a majority coalition would not pose an attack. An example of this happening in practice is DigixDAO's token holders voting to dissolve the DAO and return all ETH held in the treasury to the token holders (which was worth more than the value of all governance tokens)~\cite{yi2020digix}. However, in all other cases, where a strict subset of token holders come together to take control of a DAO, a majority coalition presents an attack.

\subsection{Human-Computer Interaction}\label{sec:HCI}

\begin{ndef}
    \emph{Human-computer interaction (HCI)} attacks aim to manipulate the voting process by exploiting user-facing interfaces and applications or human behaviors involved in the DAO's voting process.
\end{ndef}

This family of attacks lies at the boundary between the blockchains (computers) and humans. The attack vectors in this family do not exploit vulnerabilities in the underlying governance protocol itself, but rather in the interfaces, applications, or human behaviors surrounding DAO governance. 

\T{User Interface Issues (HCI1).} Many users participate in the voting process through aggregator websites that provide a convenient \textit{user interface (UI)}. Thus, bugs or malicious code in these UIs can lead to users not voting as they intended or not being able to vote at all. For example, users voting through a UI typically sign a vote transaction prepared by the UI. If this transaction is incorrectly prepared, users will potentially vote differently than they intended by signing the transaction. An incident of this type occurred with Tally, a closed-source and widely-used UI for on-chain governance.

\begin{casestudybox}{Tally}
    On 19 August 2021, a bug, which had persisted from 30 April to 19 August 2021, was discovered on Tally~\cite{2023tallybug}. The bug inadvertently altered the voting process: transactions of users wishing to vote \textit{against} a proposal were erroneously constructed by Tally. This led to these votes being recorded as votes \textit{in favor} on the blockchain. The issue went unnoticed since the transaction arguments were not presented in an easily understandable format, making it challenging for users to notice the discrepancy between their intended vote and the registered vote.
\end{casestudybox}

While there is no evidence to suggest that this bug in Tally significantly influenced the outcomes of any votes, it nonetheless highlights a critical vulnerability in centralized, closed-source front-ends for governance systems. Once a vote is cast on-chain for a proposal, it cannot be retracted or altered. This means that if a user realizes their vote has been incorrectly cast due to a platform error, they are powerless to correct it. Thus, bugs in UIs can heavily influence the voting process in DAOs, and the possibility of inserting malicious code into these UIs poses a serious risk for DAOs.

On a similar note, the unavailability of the aforementioned UIs can pose a threat to a functioning DAO governance vote. The unavailability could be caused by technical issues with the UI as well as by deliberate \textit{Denial of Service (DoS)} attacks. If widely-used UIs became unreachable ahead of a vote, users relying on these platforms to cast their votes might be deterred or prevented from voting. To the best of our knowledge, no such attack has taken place or been attempted. Nonetheless, it presents a risk worth considering for DAOs when designing their governance systems.

\T{Proposal Obfuscation (HCI2).} Obfuscation of the real intent of a proposal is a further possible attack vector, which presents a risk to DAOs, especially in combination with a weak validation of the proposal -- making sure that the proposal description matches its contents. Take as an example a proposal that appears to be a legitimate proposal but, in reality, inserts malicious code that allows the attacker to steal the DAO's funds. Such an attack was successfully performed on the Tornado Cash governance.

\begin{casestudybox}{Tornado Cash}
On 20 May 2023, an attacker gained control of the governance system of Tornado Cash~\cite{dotan2023vulnerable,behnke_explained_2023}. The attacker purchased TORN tokens through decentralized exchanges and imitated a previously accepted proposal. Due to the striking resemblance to this earlier proposal, the new, malicious one was also approved by the community. However, there was a critical and deliberate difference in the attacker's proposal: it included a self-destruction feature. After the proposal was approved, the attacker activated this self-destruction functionality, destroyed the existing proposal contract, and replaced it with malicious code. The newly inserted code allowed the attacker to withdraw TORN tokens, i.e., the DAO's governance tokens.
\end{casestudybox}

The Tornado Cash incident highlights a general vulnerability in governance mechanisms of decentralized platforms: the lack of a guaranteed match between a proposal's description and its actual code. Proposals might have unintentional errors or, as in the Tornado Cash case, be subject to deliberate manipulation. Notably, the Tornado Cash attack is not the only example of a malicious mismatch between a proposal's description and implementation. The proposal of the flash loan attack on Beanstalk (see Section~\ref{sec:acquisition}) claimed to be donating funds to Ukraine but in reality, stole the DAO's assets~\cite{beanstalk2022}. 

\T{Proposal Spam (HCI3).} A further attack vector that can be utilized to hide a malicious proposal is to spam the protocol's governance with many proposals, such that the malicious proposal is hidden in a flood of proposals. One notable example was a governance attack on Synthetify -- a protocol on the Solana blockchain whose DAO had been inactive since December 2022. The following case study details the attack, which also involved aspects of token control attacks (see Section~\ref{sec:acquisition}).

\begin{casestudybox}{Synthetify}
On 17 October 2023, an attacker gained access to the assets controlled by Synthetify's DAO~\cite{2023synthetify1,2023synthetify2,2023synthetify3}. The attacker first bought sufficient amounts of the protocol's governance token SNY to make a proposal and to hold more tokens than the three biggest holders. Then the attacker used spam to distract from the attack. In particular, the attacker created more than 20 spam proposals over two months and tested whether they would go unnoticed over the seven-day voting period. No one but the attacker voted on any of these proposals, i.e., the attacker was able to pass them without a problem. Knowing that no one was paying attention, the attacker then hid malicious code that allowed them to withdraw the funds controlled by the governance. The proposal passed without any opposition.
\end{casestudybox}

Many protocols attempt to protect against such attacks by only allowing one active proposal per account, which must sufficient tokens to exceed the proposal threshold. Nevertheless, workarounds might still pose a threat to DAOs. Consider the following workaround for DAOs utilizing the delegation model. The attacker creates a proposal with one account to which they delegate their tokens. The attacker waits for the votes to come in and cancels the proposal after a significant proportion of votes have been cast. Then, the attacker delegates the tokens to another account. The attacker then creates a new proposal and continues in this fashion in hopes of tiring the DAO's voters who pay fees for every vote.

\T{Social Infiltration (HCI4).} Individuals and institutions can take up positions of power in DAOs. For instance, delegates often vote with significantly more tokens than they hold. Moreover, some DAOs grant certain powers in the governance process to \emph{multisignature addresses (multisig)} which are jointly controlled by multiple key holders. The members of the multisig are chosen and voted upon by the DAO. One can imagine that malicious parties can maneuver themselves into these positions of power and then use their position to attack the protocol. The scandal surrounding Wonderland DAO~\cite{2022wonderland} highlights the potential risk that can stem from social infiltration. The treasury manager was found out to be Michael Patryn, a convicted criminal who had hidden his identity.

\T{ Behavioral Manipulations (HCI5).} 
Contrary to many voting systems, preliminary results of DAO votes are known to everyone. In a system where voting is associated with high costs, access to interim results could be seen as beneficial, as voters can be mobilized only if needed. However, access to preliminary results also opens up attack vectors. Yaish et al.~\cite{yaish2024strategic} highlight these attack vectors, which are attested by a large body of work on voting systems and online polls~\cite{callander2007bandwagons, zou2015doodle, morton2015exit, meir2020strategic, araujo2022casting}.

First, voters might be manipulated not to vote because they observe that their preferred outcome appears to have garnered enough support to win. An attacker can then vote at the last moment, not offering others time to react. This behavioral pattern called \textit{vote sniping} has been reported anecdotally before~\cite{evan2019aragonperil}. Rosello~\cite{rossello2024blockholders} draws parallels to corporate governance and empirically shows the negative effects vote sniping has on token value.

Conversely, attackers voting early might sway uninformed voters to follow their direction. This is commonly referred to as \textit{bandwagon voting}, an effect supported by a large amount of empirical evidence \cite{zou2015doodle, morton2015exit, araujo2022casting}. Yaish et al.~\cite{yaish2024strategic} analyze this setting theoretically, and show that interim results piled with high voting costs can entice informed voters to follow a mixed strategy of voting either early or late.

\subsection{Code \& Protocol Vulnerability}\label{sec:technical}

\begin{ndef}
    \emph{Code and protocol vulnerability (CP)} attacks exploit code or logic vulnerabilities, either in the governance smart contracts or the protocols they are connected to. 
\end{ndef}

\T{Code Vulnerability  (CP1).} To attack a DAO, an attacker can take advantage of any existing bugs in the governance smart contracts. The arguably most prominent attack on a DAO did exactly that.

\begin{casestudybox}{The DAO}
The DAO was a crowd-funded investment fund and one of the first DAOs. On 17 June 2016, an attack on The DAO occurred~\cite{2023thedao}. The attack exploited a loophole in the code, that allowed the attacker to perform a reentrancy attack to repeatedly withdraw ETH from The DAO~\cite{chainlink2022reentrancy}. Notably, the hack was so severe that it led to a highly controversial hard fork of the Ethereum blockchain. The majority of the Ethereum community decided to fork the chain to undo the hack's damages. The unaltered version of the chain continues to operate as Ethereum Classic. 

\end{casestudybox}

The DAO hack highlights the complexities of writing secure governance smart contracts. Given these complexities and the ongoing development of DAOs, code vulnerabilities appear infrequently. However, in some cases, these bugs are identified in audits and fixed before they can be exploited. For instance, in two DAOs (MakerDAO and Keep3R Network) vote tallying could be exploited~\cite{statemind2022kp3r,openzeppelin2019makerd}. In the case of Keep3r Network, the contracts permitted users to re-vote on a proposal but failed to properly subtract the user's previous vote.

Based on audits, the most well-known smart contract vulnerabilities apart from reentrancy and re-vote vulnerabilities include insufficient proposal validation and absence of transfer validation~\cite{2023mudus}. To prevent code vulnerabilities, re-using audited and time-tested code is typically seen as a good practice. However, mixing and matching code from different sources has caused at least two hacks too~\cite{halborn2021forcedao}.

\T{Protocol Vulnerability (CP2).} Vulnerabilities in the protocols associated with a DAO can extend to the DAO itself given the often intertwined nature of the two. One example of how vulnerabilities in a protocol can affect the DAO is the attack on Mango Markets.

\begin{casestudybox}{Mango Markets}
    In October 2022, Avi Heisenberg performed an attack on Mango Markets and its governance~\cite{2022maangomarket1}. Heisenberg manipulated the price oracle for MNGO, the protocol's governance token, that allowed him to take out massive loans against the protocol's treasury which the DAO controls. In doing so, Heisenberg effectively drained the treasury. He went on to create a proposal in the DAO promising to return the majority of the funds if the DAO agreed to repay the protocol's bad debt. Further, the attacker's proposal sought to ensure that the token holders could not pursue any legal action against the attacker. The attacker's proposal did not pass, but the DAO later passed an alternative proposal, leading to part of the funds being returned. The attacker, who publicly identified himself~\cite{2022maangomarket3} and infamously described the hack as a ``highly profitable trading strategy'', was later charged by the US government for his attack~\cite{2022maangomarket2}. 
\end{casestudybox}

The previously outlined incident exemplifies how the interconnectedness of a protocol and its DAO can pose a risk to the DAO. When such an intertwined nature is wished for or required, it is especially challenging to fully protect against such attacks, as complexity increases and attack vectors are likely unique to each protocol.

\section{Real-World Incidents \& Attacks}\label{sec:attacktable}

In the following, we analyze past attacks and incidents, as well as potential attacks described in audits and papers relating to DAOs. The data set in the paper includes all incidents known to us at the time of writing.\footnote{We collected the incidents by searching the web for papers, audits, news articles, blog posts, and tweets that discuss them as well as talking to experts in the field.} We further provide an up-to-date data set under the webpage \url{daoattacks.ethz.ch} and welcome readers to report any additional or new incidents.  

Table~\ref{tab:overview} lists all (theorized) incidents we analyzed. For each incident, we indicate the date and blockchain on which it occurred. Additionally, for real-world incidents, we indicate the purpose of the attack, whether it was successful, and if it was, the financial damage. Finally, we highlight which of the attack vectors introduced in the previous sections are utilized. We provide a summary for all (theorized) attacks in Appendix~\ref{app:details}.

Turning to Table~\ref{tab:overview}, we observe a relatively balanced distribution of attack vectors used in real-world incidents across the four previously introduced categories. Specifically, among the 28 attacks analyzed, 4 utilized at least one attack vector from the BR category, 14 employed TC attack vectors, 9 involved HCI attack vectors, and 9 exploit CP attack vectors.

Table~\ref{tab:overview} further summarizes critical vulnerabilities of DAOs that were uncovered in academic works, reported to the protocols, or discovered as part of audits. While attacks documented in academic papers and reports span multiple categories, those identified through audits almost exclusively belong to the CP category.

We only found a relatively small set of critical vulnerabilities identified by audits, limiting its representativeness. On the other hand, a closer examination of audits that did not uncover critical vulnerabilities reveals a similar skew towards CP attack vectors~\cite{2023mudus,2023mixbytes}. Although most DeFi protocols are primarily susceptible to CP attack vectors~\cite{zhou2023sok}, the governance aspect introduces an array of exceedingly complex attack vectors. These additional attack vectors are often less tangible to analyze and are typically not accounted for in audit processes. 

\input{attacks} 

Additionally, it is worth mentioning that a notable portion of attacks (specifically, 8 out of 28) combine multiple attack vectors. This heterogeneous nature of attacks targeting DAOs can make it challenging for DAOs to anticipate and protect against all potential attacks while, at the same time, striving to innovate and develop.

\section{Risk Factors}\label{sec:riskfactors}

Guided by our description and analysis of historical precedence cases, we identify seven risk factors that either directly or indirectly correlate with attacks on DAOs. Further, for a set of 26 DAOs on Ethereum and its Layer 2s, we empirically analyze how vulnerable these DAOs are for each of our identified risk factors in Table~\ref{tab:risk_factors}. These DAOs represent both the biggest DAOs in the Ethereum ecosystem in terms of the size of the treasuries or protocols they govern, along with smaller DAOs. This combination allows us to accurately portray the state of DAOs of all shapes and sizes. Note that we provide a brief description of our data collection in Appendix~\ref{app:datacollection}

\T{Voter Apathy (RF1).} If token holders do not delegate or vote themselves, it becomes much easier for an attacker to pass malicious proposals. In all but four of the DAOs we empirically analyzed in Table~\ref{tab:risk_factors}, tokens must be delegated before voting. Importantly, when voting takes place, no more delegation is possible. We show the percentage of both delegated tokens voting and the percentage of the total token supply voting on average in the last five votes -- a measure of voter apathy. Note that any tokens that are not delegated ahead of the voting period are completely excluded from voting. When regarding the first two columns in Table~\ref{tab:risk_factors}, notice the relatively low participation from the delegated tokens at 34\% across the 20 DAOs that require delegation in our data set. While some DAOs have a high participation of more than 81.13\% (i.e., Ampleforth), in other DAOs the participation of delegated tokens sits around 1\% (i.e., Pooh). Additionally, even more startlingly, of the entire token supply, on average only 5\% of tokens participate in the governance across the DAOs we analyzed.  We highlight that these low participation rates of (delegated) tokens can be seen as a considerable risk factor, as an attacker can attempt a majority attack, even when holding just a fraction of the tokens. 

\input{risks}

\T{High Governance Token Liquidity (RF2).} High governance token liquidity entails the possibility and comparatively low cost of buying or lending the governance token -- making the attack vectors in the token control category we presented in Section~\ref{sec:acquisition} feasible. Table~\ref{tab:risk_factors} shows that available liquidity on Uniswap V2 and V3 (the two biggest decentralized exchanges on Ethereum in terms of \textit{total value locked (TVL)}~\cite{dextvl}). We show the available liquidity as a percentage of (1) the proposal threshold, i.e., the minimum number of tokens required to create a proposal in the governance, (2) the delegated votes, i.e., the number of tokens required to almost guarantee success in the analyzed DAOs, and (3) the average number of tokens voting in the last five governance votes. We observe that for 17 DAOs the available liquidity exceeds the proposal threshold, whereas only for zero and two DAOs the available liquidity exceeds the delegated votes and the average number of votes respectively. While this appears promising, we underline that the figures presented are a strict lower bound as they for example do not include centralized exchanges where the available liquidity is not easily quantifiable. Even though for the analyzed DAOs liquidity currently appears low, we presented 14 attacks that still attempt to exploit DAOs through token control (see Section~\ref{sec:attacktable}). Thus, we reiterate that for a DAO's safety, lower liquidity is advantageous.

\begin{figure*}[bt]\vspace{-4pt}
    \centering
    \begin{subfigure}{0.49\linewidth}
        \includegraphics[scale =1,right]{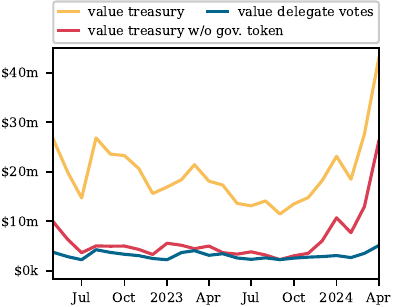}
        \caption{Ampleforth}
    \end{subfigure}
    \hfill
    \begin{subfigure}{0.49\linewidth}
        \includegraphics[scale =1,right]{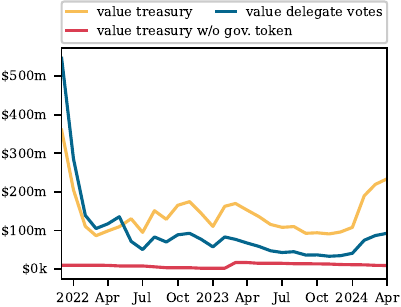}
        \caption{ENS}
    \end{subfigure}
    \\
    \begin{subfigure}{0.49\linewidth}
        \includegraphics[scale =1,right]{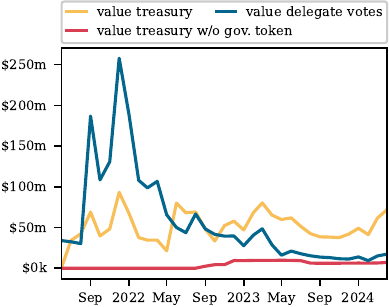}
        \caption{Gitcoin}
    \end{subfigure}\hfill
    \begin{subfigure}{0.49\linewidth}
        \includegraphics[scale =1,right]{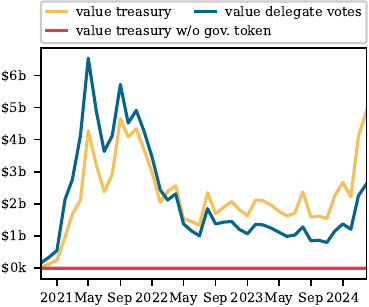}
        \caption{Uniswap}
    \end{subfigure}\vspace{-8pt}
    \caption{Comparison of treasury values and the total value of all delegated governance tokens. If the value of the treasury (yellow line) or even the value of the treasury without the governance token (red line) exceeds the value of delegate votes (blue line) this represents an economic risk.}
    \label{fig:treasuries}\vspace{-10pt}
\end{figure*}

\T{Large Treasury (RF3).} The impact and attractiveness of an attack increase, the more value is stored in the treasury. Since in the aftermath of an attack, token prices are expected to plummet, we wager that the treasury value excluding the governance token itself is the most important driving factor. Our empirical analysis in Table~\ref{tab:risk_factors} presents the treasury value with respect to the value of all delegated tokens, both with and without the governance token. A considerable chunk of DAOs (i.e., 6) hold less than 10\% of their treasury value in tokens other than their governance token and are thus likely less at risk for a governance attack that aims to empty the treasury. Startlingly, for the Ampleforth DAO, the value of the treasury without the governance tokens exceeds the value of all delegated tokens -- making it an attractive target for attacks. Additionally, we highlight that if the value of the treasury (without the governance token) exceeds 50\% of the delegated votes, 51\% attacks of token holders that have delegated their tokens could be rational. A few DAOs are close to reaching this threshold (e.g., Gitcoin) or have been in the past. Note that we are not aware of any precedence for such an attack, but protocols have forked before~\cite{2023nouns}. In addition to the empirical snapshot of 31 January 2024  presented in Table~\ref{tab:risk_factors}, for a smaller subset of DAOs we also visualize the historical value of the treasury in comparison to the delegated token values (see Figure \ref{fig:treasuries}). We observe significant fluctuations over time in the relative value of the delegate tokens in comparison to the treasury for the three DAOs: Ampleforth, ENS, Gitcoin, and Uniswap. While initially for three of the DAOs (i.e., ENS, Gitcoin, and Uniswap) the value of the delegate votes (blue line) exceeded the value of the treasury (yellow line) this is no longer the case for all of them. For these DAOs, except for Uniswap (which does not hold tokens other than its governance token), the difference between the value of the delegate votes and the value of the treasure without the governance token is shrinking over time. Finally, for Ampleforth the value of the delegate votes never exceeded the value of the treasury and also currently does not exceed the value of the treasury without the governance token. We conclude that DAOs need to constantly monitor the value of the treasury to ensure that they are not an attractive target for token control attacks.

\T{Inadequate Configuration (RF4).} Inadequate configuration of voting contracts can leave a wide scope of vulnerabilities open. We discuss the most important parameters in the following. First, \textit{proposal delay}, i.e., the delay between proposal creation and the start of the voting period, must be larger than 0 to avoid flash loan attacks. A proposal delay of 1 block, as used by DAOs (see Table~\ref{tab:risk_factors}) is also not without issues though, especially for DAOs that require delegation. Such a small delay does not leave time for non-delegated tokens to be delegated in case of a malicious proposal. For similar reasons, a short \textit{voting window}, might also be dangerous, as delegates might not be reached in time to vote against a malicious proposal. However, all DAOs we analyzed have a voting window that runs for a couple of days (e.g., there are around 7,000 blocks a day on Ethereum). Finally, adjusting the duration a proposal must remain in the timelock can also be beneficial, i.e., \textit{timelock delay}. Extending this period forces an attacker to maintain a number of votes, at least equal to the proposal threshold, for a longer duration. This approach increases the risk for the attacker and makes the potential profits less predictable.

\begin{figure}[h]\vspace{-4pt}
    \centering
    \includegraphics[scale =1]{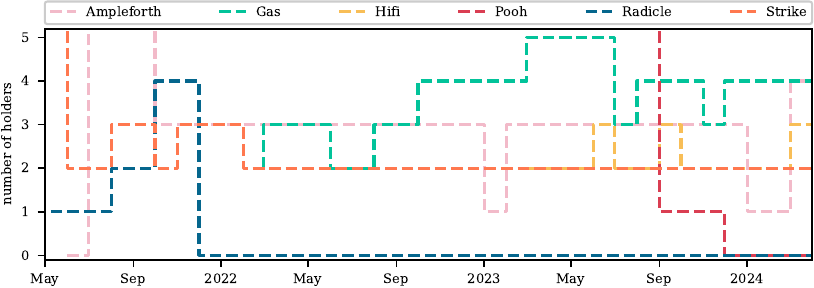}\vspace{-3pt}
    \caption{Number of holders, i.e. EAOs, who hold more tokens than delegated governance votes on a monthly basis. These holders would have the majority of the delegated votes after they delegate their tokens.}
    \label{fig:influence_centralised}\vspace{-10pt}
\end{figure}

\T{Centralization (RF5).} If a large (delegated) token supply is held only by a few addresses or entities, many attack vectors become more likely to succeed (e.g., majority coalition, whale activation). In Table~\ref{tab:risk_factors}, we show the Nakamoto coefficient of the delegate votes and the token supply, i.e., the minimum number of addresses collectively holding more than 50\% of the delegate votes and the token supply. The lower the Nakamoto coefficient, the higher the centralization. We find that, startlingly, for three DAOs the Nakamoto coefficient of the delegate tokens is one -- one delegate has the majority of delegate votes. Finally, we also consider the number of \textit{externally owned addresses (EOAs)} that hold more governance tokens than are currently delegated. Importantly, more than one holder can hold more votes than delegated governance votes, as not all tokens are delegated. These EOAs could delegate their tokens and would have a majority of the delegate tokens. In combination with a small proposal delay (RF4), they could easily acquire the majority of votes. For six DAOs there was at least one EOA that could perform such a 51\% attack on 31 March 2024. Additionally, we also analyze how this figure evolves over time for the five DAOs of these DAOs on the Ethereum blockchain in Figure~\ref{fig:influence_centralised}. We observe that for these five DAOs, there was generally at least one EOA that held sufficient tokens for a 51\% attack and thereby posed a threat. 

Table~\ref{tab:risk_factors} also shows that some DAOs have a guardian in their governance contract or in their timelock contract. This involves special rights that, for example, enable an EOA or a multi-signature wallet to cancel proposals. On the one hand, this functionality can be abused and lead to a situation where only decisions that aren't canceled by the guardian can be made, or if not implemented carefully, give the guardian privileged access to the treasury or other critical infrastructure. On the other hand, a trustworthy guardian can mitigate the effect of malicious proposals.

\T{Code Uncertainties (RF6).}
When smart contracts are created, the contract bytecode that is uploaded on-chain can contain arbitrary logic. As smart contracts may contain various unknown mechanisms, any uncertainty can be viewed as risky. Firstly, the smart contract creators should thus publish the source code, allowing anyone to verify its logic.
Additionally, some code functionalities are associated with a higher risk. For instance, the presence of a mint functionality might allow an attacker to create more tokens. The mint function can be a particular risk as it allows attackers to empty the liquidity pools with the governance token (see Build Finance and Curio in Appendix~\ref{app:details}). 
We observe in Table~\ref{tab:risk_factors} that five of the analyzed DAOs implement such functionality in their smart contract. Risks are also associated when external calls are allowed, and when a proxy contract is used (as the proxy contract may be changed to point to a different contract, bypassing the DAO)~\cite{consensys2023bestpractices,boringsecurity2023proxy}. Table~\ref{tab:risk_factors} shows that only for one DAO the ownership of the contract was renounced. This is considered a good practice, as the contract then cannot be called with elevated owner privilege anymore~\cite{certik2023general}.

\T{Lack of Reliable Communication Channels (RF7).}  DAO community members mainly communicate through X (formerly Twitter), Telegram, and Discord. These platforms are crucial parts in defending an attack, as seen in the Indexed Finance case study presented in Section~\ref{sec:acquisition}. Still, it is difficult to reach all delegates and token holders, especially if the projects are no longer active, as was the case for Indexed Finance. Thus, better infrastructure to reliably reach holders and inform them about ongoing governance votes would be beneficial. To the best of our knowledge, none of the DAOs we have analyzed have implemented any more reliable communication channels than those mentioned in the beginning.

\vspace{6pt}\noindent
The described risk factors are diverse and thus preventing against them all simultaneously is a difficult task. Our empirical evaluation of 26 DAOs and their susceptibility to these risk factors also revealed that smaller DAOs tend to be more at risk. For these DAOs, in the absence of the same resources as their larger counterparts, it is likely especially hard to protect against all possible attack vectors. Thus, especially smaller projects should weigh the benefits and disadvantages of a DAO carefully. For those, that choose a DAO as their governance form we continue by describing and discussing safeguards.

\section{Mitigation and Safeguards}\label{sec:safe}

We present and discuss mitigation strategies to reduce risks. Throughout, we distinguish between mitigations that lower the impact (\faRocket) of an attack and those that lower the likelihood (\faQuestionCircle) of success for an attack. In parentheses, we specify which attack vector categories are targeted.

\T{Conservative Implementation \faRocket~\faQuestionCircle~(BR, TC, HCI, CP).} Through conservative implementation, DAOs make sure that exogenous factors cannot be exploited to attack a DAO. Examples include \textit{limiting the number of proposals} that can be made by a single proposer at any given time~\cite{uniswapsourcecode} to prevent spamming attacks and having long enough \textit{proposal delays} (see Section~\ref{sec:riskfactors}). This involves trade-offs, as extending the on-chain proposal process can make an attack less appealing, but it also slows down governance in general. Thus, a balance must be struck between a DAO's agility and safeguarding against potential attacks. We further note that a lack of agility for a DAO can pose additional risks depending on the protocols they govern~\cite{Heimbach2023defi,Heimbach2023short}.

\T{Limiting the Governance Scope \faRocket~(BR, TC, HCI, CP).} Another approach to lessening the impact of attacks is for a DAO to add restrictions on its action space that can reduce the attack surface. For instance, if the DAO is only granted control over a few parameters, the extent of potential attacks is much narrower. Additionally, one can imagine only allowing a proposal to spend a fixed maximum amount of the treasury. 

\T{Emergency Shutdown \faRocket~(BR, TC, HCI, CP).} Implementing an emergency shutdown is a very invasive mitigation strategy. Here, a set of holders can halt all transactions. In the case of MakerDAO~\cite{makerdao2023shutdown}, the emergency shutdown allows token holders to receive a share of the treasury, mitigating potential attacks that were underway.

\T{Governance Forks \faRocket~\faQuestionCircle~(BR, TC, HCI, CP).} A similar, but less drastic, safeguard could be achieved through forking, a design primitive where a fraction of token holders can vote to create a fork of the DAO. For instance, The DAO allowed token holders to create \textit{child DAOs} and later withdraw their portion of the DAO deposits from there. Another example of the occurrence of a DAO fork is NounsDAO: A large fraction of holders decided to leave the original DAO for a forked DAO taking with them their proportional share of the treasury~\cite{defiant2023nouns}. The forked DAO then allowed each token holder to \textit{rage-quit} and retrieve their individual share of the treasury. This process is usually not very fast, and thus can typically only prevent foreseeable attacks. Nonetheless, allowing DAOs to fork is a possibility to prevent a majority (coalition) from taking over a DAO (and its treasury). With a fork, a minority would still have the possibility to take their part of the DAO's assets. However, if a DAO governs more than a fungible treasury, e.g., the parameters of a lending protocol, forking may of course not be a viable option.

\T{User Authentication \faQuestionCircle~(BR, TC).} Through user authentication, voting power is to be constrained on a per-person basis. This can enable different voting mechanisms, that might be less vulnerable to token control attacks, such as \textit{quadratic voting} (voting power is proportional to the square root of tokens owned) and \textit{democratic voting} (one person one vote). Examples of user authentication include \textit{know-your-customer (KYC)} or decentralized identifiers, like \textit{Proof-of-Personhood}~\cite{borge2017pop}. The Optimism Governance recently implemented a form of user authentication. In particular, they implement a bicameral governance design, with a token house~\cite{2022optimismtoken} (one token one vote) and a citizen house~\cite{2022optimismcitizen} (one person one vote), only those with citizenship can vote in the citizen house.

\T{Ballot Privacy \faQuestionCircle~(HCI)} Ensuring ballot/tally privacy during the voting period can help in mitigating behavioral manipulations (HCI5). Cicada~\cite{glaeser2023cicada} is an existing framework for the EVM which achieves this. While it is costly to implement on Ethereum, the costs are more reasonable on L2s.

\T{Governance Tools \faQuestionCircle~(HCI).} The development of novel governance tools reduces the hurdles of participation in governance and can help prevent HCI attacks. For instance, through better communication and notifications on current proposals, voter apathy can be combated. Moreover, they may provide the necessary education for voters to be able to make informed decisions more easily, also mitigating behavioral manipulations (HCI5). We believe it is important that these tools are open-source (i.e., such that bugs as in the Tally~\cite{2023tallybug} case are less likely to happen) and that they cannot easily be spammed or taken down. While these tools can do a great part in reducing the load in governance participation, they can become potential attack victims themselves (see Section~\ref{sec:HCI}).

\T{Veto Power \faQuestionCircle~(HCI).} DAOs may also introduce a veto functionality. Through a veto, a small group of holders can prolong a vote, giving the holders more time to counter malicious proposals. Excessive use of veto power itself leads to issues, but we hypothesize that incentives could deter its misuse (e.g., vetoing could be made expensive).

\T{Objection Phase and Vote Extension \faQuestionCircle~(HCI).} A more targeted safeguard consists of the addition of a second round of voting (also called \textit{objection phase}), where voters can only vote against the proposal, or change their vote from in favor to against. This has been introduced by Lido \cite{lido2022twophase}, with the goal to protect the DAO from vote sniping (see HCI5). Other proposed remedies to vote sniping include the extension of the voting period after high activity (or sway votes) are observed, as well as randomized voting durations. Both have recently been suggested by Decentraland DAO~\cite{decentraland2024lastminute}.

\T{Scheduled Voting Windows \faRocket~(HCI).} Some protocols are experimenting with votes being scheduled on a regular basis (e.g., once a month). This can prevent proposal spam (HCI3) to reduce voter apathy and dampen the effect of behavioral manipulations (HCI5).
\T{Escape Hatches \faRocket~(CP).} Escape hatches can be added to DAOs to limit the severity of an attack. The \textit{Decentralized Escape Hatch} proposed by Eyal and Sirer~\cite{eyal2016decentralized} for example suggests that outgoing transactions can be buffered (e.g., for 24 hours). Buffered transactions can then be reversed automatically, by specifying programmatic invariants. Such invariants could for example limit the outflow over time, or check whether outflow is consistent with respective inflows. Note that invariants themselves are hard to get right. The authors, thus, also suggest community involvement by crowdsourcing the reversal, for example through a majority involvement.

\T{Audits \faQuestionCircle~(CP).} Last but not least, audits by external companies can help verify that the DAOs underlying smart contracts are implemented to the state-of-the-art. Audits will make sure that code best practices are respected~\cite{2022bestpractices}, according to the platform and language used. We observe that audits typically focus on technical vulnerabilities. While we find that they could also consider the more governance-specific attack vectors we present, technical audits also hold immense importance for the security of DAOs.

\section{Related Work}

Possible attacks on DAOs have been discussed in blog posts almost as long as DAOs have existed~\cite{mark2016thedao, daian2018dark} including by Ethereum's founder Vitalik Buterin~\cite{buterin2017voting, buterin2021voting}.
Among other things, they discuss the risks of low voter participation, centralization, game-theoretic attacks, and vote buying, as well as possible mitigation strategies such as \textit{limited governance}, \textit{non-coin-driven governance}, and \textit{skin in the game}.

An early instance of a DAO governance attack documented in academic literature is a potential attack on the governance of the MakerDAO protocol, the centerpiece of DeFi at the time, by Gudgeon et al.~\cite{gudgeon2020crisis}. More recently, Augsten et al.~\cite{austgen2023dark} have discussed the potential of hidden vote buying in DAO governance facilitated by smart contracts, i.e., what is referred to as \textit{Dark DAOs}. Related to the attack on DAO governance, the term \textit{Governance Extractable Value (GEV)} has been coined to describe the potential value that can be gained from influencing DAO governance votes~\cite{lee2021gev}. Note that the term is an homage to the widely-studied concept of \textit{Miner/Maximal Extractable Value (MEV)}~\cite{daian2020mev}.

Two recent systematizations of knowledge (SoKs) cover topics related to DAO attacks: Zhou et al.~\cite{zhou2023sok} survey hacks and incidents in DeFi protocols in general. However, most described attacks are not attacks on the protocol's governance system, which we focus on in this paper. A general overview and systematization of the concept of governance for blockchains and blockchain-based protocols can be found in the SoK by Kiayias and Lazos~\cite{kiayias2023sok}. It discusses the governance processes of blockchains such as Bitcoin and Ethereum, along with examples of protocols running on blockchains -- which are the focus of our SoK. Additionally, Ethereum's governance process, including which actors have how much influence on it, has also been studied in detail by Fracassi et al.~\cite{fracassi2024decentralized}. To the best of our knowledge, this paper represents the first SoK to study attack vectors, risks, and possible mitigation of attacks on the governance of DAOs.

Recently, the literature surrounding DAOs has rapidly expanded, including two reports of the WEF on DAOs~\cite{wef2022daos, wef2023daos}. This encompasses a flurry of empirical studies on a variety of DAOs covering aspects such as token distributions, voting turnout and voting behavior~\cite{fritsch2022analyzing, barbereau2022distribution, feichtinger2023shortcomings, barbereau2023distribution, sun2023decentralization, dotan2023vulnerable, sharma2023unpacking, kitzler2023governance, messias2024understanding}.
In particular, many of the studies (see e.g., Feichtinger et al.~\cite{feichtinger2023shortcomings}) make a number of observations relevant to attacks covered in this paper: They reveal that a majority of voting power is often concentrated in the hands of a very small number of holders and delegates. Additionally, they highlight that participation rates in governance votes are frequently low across many DAOs.

The vast majority of DAOs today, including those covered in the aforementioned studies, use simple token voting (one-token-one-vote). An alternative governance model using \textit{vote escrowed} tokens (governance tokens locked for a fixed time period), which is for instance used by Curve and Balancer, is discussed by Lloyd et al.~\cite{lloyd2023veToken}.

Finally, Tan et al.~\cite{tan2023open} describe open research problems surrounding DAOs in fields ranging from computer science and economics to ethics, law, and politics.

\section{Conclusion}

In this paper, we systematically analyzed potential attacks on DAOs along with 28 real-world incidents to illustrate the scope of security vulnerabilities. By describing and categorizing the multitude of attack vectors, we provided a comprehensive overview of the threats faced by DAOs. Additionally, we identified and empirically measured risk factors across a set of 26 DAOs, offering insights into the prevalent risks and their impact.

We believe that it is highly advisable for a DAO to engage early with the possibility of such an attack, to monitor parameters closely, and to ensure that an attack does not become economically attractive. Understanding these challenges is critical when designing and operating a DAO, and poses a significant challenge to DAOs. Ultimately, with our systematization of attacks on DAOs, the vulnerabilities of DAOs, and possible safeguards, we seek to arm future DAO designs with the necessary knowledge to anticipate and mitigate these threats.



\bibliography{bibliography}

\appendix

\newpage
\section{Analyzed Incidents, Attacks \& Audits}\label{app:details}
In the following, we provide a brief description of each of the attacks and incidents analyzed in Table~\ref{tab:overview}. Note that we present the incidents and attacks in alphabetical order. Additionally, for those already discussed in case studies, we refer to the section where the case study can be found. 

\T{Audius (Jul 2022)~\cite{cointelegraph2022audius}}
A bug in the initialization contract was exploited by an attacker to drain \$6.1M worth of Audius governance tokens from the DAO's treasury and gain control over the DAO. An assembled response team was able to use the same vulnerability to gain back control over the DAO within a few hours and patch the contracts~\cite{audius2022postmortem}.

\T{Agora Audit (May 2023)~\cite{obront2023agora}} The audit by an independent security researcher brought to light that certain proposals could be passed irrespective of whether votes in favor of the proposal were more numerous than votes against it. Instead, the contract only checked that a given quorum of yes votes was reached (As is customary in DAOs, the required quorum was set much lower than 50\% of voting power).

\T{Bandwagon Voting (Feb 2024)~\cite{yaish2024strategic}} Yaish et al.~\cite{yaish2024strategic} point to empirical evidence from national elections and online voting (such as Doodle polls) with public interim results, showing that late voters are more likely to vote for leading candidates. They argue that this so-called bandwagon effect might also occur in DAO governance. To this end, they propose a theoretical model capturing the DAO voting process. Based on this model, they are able to show that early voting is indeed part of the rational strategy.

\T{Beanstalk (Apr 2022)} See Section~\ref{sec:acquisition}.

\T{BigCap DAO (Sep 2023)~\cite{2023bigcap}} A malicious proposal on the BigCap DAO, which was ultimately rejected, tried to steal the treasury. The proposal copied a previous proposal and the attacker had previously acquired the BIGCAP tokens, i.e., the DAO's governance tokens, through a decentralized exchange~\cite{2023bigcapetherscan}.

\T{Binance (Oct 2022)~\cite{2022binance}} UNI tokens (Uniswap governance token) deposited with Binance were delegated, even though the cryptocurrency exchange says that it does not vote with its users' tokens. Binance later commented on the incident and said that it was an accident.

\T{Build Finance (Feb 2022)~\cite{2022buildfinance1,2022buildfinance2,2022buildfinance3,2022buildfinance4}} A first attempt at a takeover of the DAO was made on 9 February 2022 with a proposal that would allow the attacker to mint the DAO's governance token -- BUILD. This attempt was picked up in the DAO's Discord channel, where voters were urged to vote against the proposal and subsequently failed. On the next day, the attacker tried their luck again, transferred the tokens to a different wallet, and created a second proposal. This proposal went undiscovered, passed in favor of the attacker, and gave the attacker control of the DAO. The attacker went on to mint BUILD tokens and emptied the liquidity pools that held BUILD on various decentralized exchanges~\cite{2022buildfinance5}. The attacker received the equivalent of \$470,000. Note that some of the BUILD tokens were bought on decentralized exchanges.

\T{Constitution DAO (Jan 2022)~\cite{hacken2022constitution}} An audit found that the project owners (that control a multisig), can replace the management contract, allowing them to move funds, and change the project logic arbitrarily.

\T{Compound (Feb 2022)~\cite{2022compound}} An address proposed to add TUSD as a collateral asset on Compound, the address had previously received COMP (Compound's governance token) tokens worth \$9M from Binance. As Justin Sun had previously borrowed COMP tokens and sent them to Binance, it is alleged that he is behind this governance attack which ultimately failed.

\T{Curio (Mar 2024)~\cite{halborn2024curio}} An attacker minted approximately 1B CGT, Curio's governance token. The primary vulnerability was a flaw in the voting power privilege access control. The attacker acquired a small amount of tokens, and after elevating its voting power, could approve a malicious contract as an external library. By calling this external library through a \textit{delegatecall}, that attacker gained the capacity to perform various nefarious activities, such as token minting.

\T{Curve A (ongoing)~\cite{2023curvewars}} The Curve Wars is an ongoing competition between various DeFi projects to attract rewards to their respective liquidity pools. Protocols bribe veCRV, i.e., Curve's governance token, token holders to vote to distribute rewards towards their pools on Curve.

\T{Curve B (Nov 2021)~\cite{2022curveb1,2022curveb2}} Mochi Inu set up a pool on Curve pool that attracted significant liquidity, they were able to push up the rewards of the Mochi Inu pool and gain an outsized influence in the Curve governance. Even though such practices are constantly ongoing on Curve (see Curve A), this particular attack was stopped by Curve's Emergency DAO, which cut off the pool's rewards. 

\T{Curve C (Jul 2020)~\cite{trailofbits2020curve}} An audit found that Curve's DAO was using a voting contract (Aragon) with known issues. Next to code vulnerabilities that allowed users to abuse their voting power, the auditors also pointed out the dangers of a potential spam attack.

\T{DAO Maker (Mar 2021)~\cite{hacken2021daomaker}} An audit found that contract owners have elevated privilege access to certain functionality, allowing them for example to drain user funds. 

\T{Dark DAOs (Jul 2018)~\cite{daian2018dark,austgen2023dark,augsten2024dark}} Dark DAOs are DAOs whose general goal it is to ``subvert credentials in an identity system''~\cite{austgen2023dark}. More concretely, the recently proposed prototype is designed to facilitate vote-buying for DAOs on Ethereum, by offering varying levels of confidentiality to its participants~\cite{austgen2023dark}.

\T{Force DAO (Apr 2021)~\cite{halborn2021forcedao}} The decentralized hedge fund was built on smart contracts that were written using code from two different sources. Due to their mismatch in error handling, by performing transfers that failed, attackers were able to obtain tokens that granted them access to shares of the vault for free. 

\T{GameDAO (Aug 2021)~\cite{certik2021gamedao} } An audit by Certik uncovered that an address has privileged ownership over many contract functionalities, giving it control over the minting and burning of NFTs to and from any account.

\T{Genesis Alpha DAO (Feb 2019)~\cite{levi2019genesisalpha}} The Genesis Alpha DAO was an experimental project led by the DAOstack initiative (shutdown in late 2022), that proposed a suite of governance solutions. The Genesis Alpha contract was a DAO built on top of the Arc platform. Arc was capable of supporting multiple DAOs through the same voting contracts, which DAOstack argued made created DAOs more secure, as all the individual pieces of the platform could be audited separately~\cite{levi2018arc}. On 5 February 2019, however, attackers showed that the lack of separation of code was exploitable, as they were able to drain the Genesis Alpha treasury, through another DAO. More precisely, due to a bug in the code, the attackers were able to assign themselves voting rights for the Genesis Alpha DAO. In the same atomic transaction, the attacker then created, voted for, and passed a proposal that transferred the contents of the Genesis Alpha DAO treasury (around \$15,000 in ETH and GEN tokens) to himself. The Arc Platform had previously been audited. Speculation arose that the attack might have happened since a bounty for a newer DAO contract had recently been put up. DAOstack wrote that this showed that \textit{community bug bounties} work~\cite{levi2019genesisalpha}.

\T{Hoprnet Token (Jun 2021)~\cite{chainsecurity2021hopr}} An audit pointed out that the function allowing users to claim governance tokens used an incorrect assertion. This would have allowed users to retrieve more tokens than their allocated amount. The assertion was rewritten.

\T{Indexed Finance (Nov 2023)~\cite{2023indexed}} See Section~\ref{sec:acquisition}.

\T{Kleros (Dec 2023)~\cite{kleros2023governor}} Kleros is a decentralized arbitration protocol whose DAO was attacked in December 2023. In this DAO, voting happens off-chain, and only proposals that are to be implemented are submitted on-chain. The attacker deployed a custom smart contract to submit the proposal list to the Governor smart contract. Most likely, a custom smart contract was used to evade the notification systems. The attacker's proposal would have transferred 46 ETH out of the DAO if it had been successful. However, the attack was discovered and stopped by issuing a counter-proposal, which was selected instead of the malicious proposal by Kleros' Court system.

\T{KP3R Network (Sep 2022)~\cite{statemind2022kp3r}}  An audit discovered that the contracts permitted users to re-vote on a proposal but failed to properly subtract the user's previous vote. The vulnerability went unnoticed for two years.

\T{MakerDAO A (Feb 2020)~\cite{gudgeon2020crisis}}
A prevented attack on the MakerDAO stablecoin protocol was disclosed to MakerDAO by the researchers who discovered it. If successful, the attack would have potentially allowed an attacker to steal \$0.5b worth of MKR and DAI collateral, as well as minting an unlimited supply of DAI tokens. The attacker simply needed to obtain a sufficient amount of MKR tokens to win a vote (see \emph{Token Purchase}, \emph{Token Loan}, \emph{Flash Loan} in Section~\ref{sec:acquisition}). In particular, using flashloaned tokens to vote was possible at the time.

\T{MakerDAO B (Oct 2020)~\cite{2020makerdaob1}} In an effort to speed up the passing of a proposal, BProtocol first took out a flashloan on dYdX of ETH which they deposited on Aave to borrow MKR tokens, i.e., MakerDAO's governance token. These MKR tokens were then used to vote in favor of the proposal and schedule the execution of the proposal. In the aftermath, additional safeguards were implemented by the DAO~\cite{2020makerdaob}.

\T{MakerDAO C (Jan 2022)~\cite{2022makerdaoc}} Justin Sun allegedly borrowed large amounts of MKR, the governance token for the MakerDAO, to vote in a poll to create a TUSD-DAI peg stability module. He, however, returned the tokens after his actions were noticed and did not end up waiting. 

\T{MakerDAO D (May 2019)~\cite{openzeppelin2019makerd}} An audit discovered that the contract logic permitted an attacker to remove votes of other users from proposals, as well as indefinitely lock other users’ tokens on any given proposal.

\T{Mango Markets (Oct 2022)} See Section~\ref{sec:technical}.

\T{Nexus Mutal (Feb 2020)~\cite{2020nexusmutal}} The Nexus Mutual team was informed that the DAO's advisory board could whitelist a proposal, but that a different proposal would be executed in practice. In more detail, a proposal to upgrade the protocol would be whitelisted and then replaced with a malicious proposal that would execute.

\T{Paladin Lending (ongoing)} See Section~\ref{sec:bribe}.

\T{POA Network (Sep 2018)~\cite{chainsecurity2018poa}} A security issue of high severity was uncovered in the protocol's governance contract. The audit found that a three owner addresses have the power to manipulate the vote outcome contrary to the majority opinion. The issue was successfully mitigated before deployment of the contract.

\T{Snapshot X (Jul 2023)~\cite{chainsecurity2023snapshotx}} An audit pointed out that the Snapshot X on-chain voting protocol computed the voting power in an ``unfair'' way. When a proposal was submitted, the voting power of voters was set according to the token distribution at the block preceding the proposal creation. This gave the proposer over-sized power, as they were in the unique position to benefit from this knowledge, e.g., by buying or lending tokens for this single block. Moreover, other voters did not get that opportunity, as they were unaware of the incoming proposal. In the remedied version, the voting power is determined one block before the start of the voting period instead, which can be set sufficiently far in the future, for other voters to also ``prepare'' for the vote.

\T{Steemit (Feb 2020)~\cite{2020steemit,garimidi2023dao}} Steemit is a protocol whose on-chain governance is controlled by 20 witnesses selected by the STEEM token holders, i.e., the protocol's native and governance token. Justin Sun acquired 30\% of the STEEM token supply, which was noticed by the protocol's witnesses. The witnesses went on to freeze the tokens bought by Justin Sun. A back and forth between Justin Sun and the protocol followed, in a quest to install their preferred 20 witnesses to control the protocol. Eventually, Justin Sun came out victorious and gained control of the protocol. 

\T{Synthetify (Oct 2023)} See Section~\ref{sec:HCI}.

\T{The DAO (Jun 2016)} See Section~\ref{sec:technical}.

\T{Tally (Apr 2021)} See Section~\ref{sec:HCI}.

\T{Temple DAO (Oct 2022)~\cite{2022temple1,2022temple2,2022temple3}} An attacker took advantage of insufficient access control to the migrateStake function in the Temple DAO smart contract. In more detail, anyone was able to call the migrateStake functions and there was no validation on the oldStaking parameter. This allowed the attacker to insert a malicious contract as the oldStaking parameter and insert an arbitrary amount, which was then withdrawn from the DAO and received by the attacker. The attacker drained around \$2.4M from the DAO. 

\T{Tornado Cash (May 2023)} See Section~\ref{sec:HCI}.

\T{True Seigniorage Dollar (Mar 2021)~\cite{2021true1,2021true2}} The attacker first acquired a large amount of TSD tokens, i.e., the protocol's governance tokens. With these tokens, the attacker went on to propose and vote on a proposal that would replace the token contract with malicious code. The attacker then minted TSD tokens and swapped these for BUSD (a stablecoin pegged to the \$) on decentralized exchanges. 

\T{Venus (Sep 2021)~\cite{mandal2021venus}} 
The DeFi protocol Venus experienced a takeover attempt. A proposal (VIP-42~\cite{venus2021vip42}) suggested that a new collective (``Team Bravo'') should be created to take over voting and financial authority over the protocol. The proposal's author promised a large sum of governance tokens (\$29M) to be paid out to any supporters. Although this bribing scheme seemed to work (the proposal was initially passed), the contract developers unilaterally canceled the proposal. This also ascertained the oversized influence held by the original contract developers.

\T{Vote Sniping (May 2024)~\cite{rossello2024blockholders}} Rossello~\cite{rossello2024blockholders} provides empirical evidence that vote sniping, i.e., changing the outcome of the vote in the last moments, occurs frequently, and has negative effects on the governance token price.

\T{Wonderland DAO (Jan 2022)~\cite{2022wonderland}} Wonderland DAO discovered that Michael Patryn was their treasury manager. He had hidden his identity behind the pseudonym 0xSifu. Possibly, he hid his identity as he is the co-founder of the failed cryptocurrency exchange QuadrigaCX and a convicted criminal. 

\T{Yam Finance (Jul 2022)~\cite{decrypt2022yam}}
An attacker created a malicious proposal that would have given them control over the treasury worth \$3.1M at the time. The true intention was hidden behind a bogus proposal text. After being notified of suspicious activity, the team behind Yam Finance used their privileges to cancel the proposal. Interestingly, a few days prior, the team was faced with another difficult situation, as on the off-chain platform Snapshot, the token holders had voted to allow governance tokens to be redeemed for a pro-rata share of the treasury (tokens were trading at \$0.11, while they could have been redeemed for \$0.25). The Yam Finance team called for a re-vote. However, a few months later an on-chain proposal passed, which allowed tokens to be redeemed~\cite{twitter2023yam}.

\T{Yuan Finance (Sep 2021)~\cite{yuan2021takeover}}
An attacker bought 4\% of the governance token supply and used this voting power to pass a proposal to change the protocol's governance to a contract under their control. Through this new-gained control, the attacker acquired the ability to mint an essentially unlimited number of tokens, which they used to drain exchange pools. Notably, subsequent attempts at draining the DAO's treasury were thwarted, as the original contract deployers had full veto power over proposals. To the best of our knowledge, in total, the attacker took away 63.8 ETH (worth ~\$219K at the time) and 63.1K USDx (worth ~\$62.7K at the time)~\cite{etherscan2021yuan}.

\section{Data Collection}\label{app:datacollection}
We run an erigon Ethereum client (connected to a lighthouse client), a geth Arbitrum client, and a geth Optimism client to collect first hand on-chain data. From the established databases we extract all metrics used, such as governance participation, token distribution, and available liquidity. The Python code used for the data analysis is provided in the following repository~\cite{anonymous2024git}.

\section{Paladin Lending}\label{app:paladin}

\begin{figure}[ht]
    \centering
    \includegraphics[width=\columnwidth]{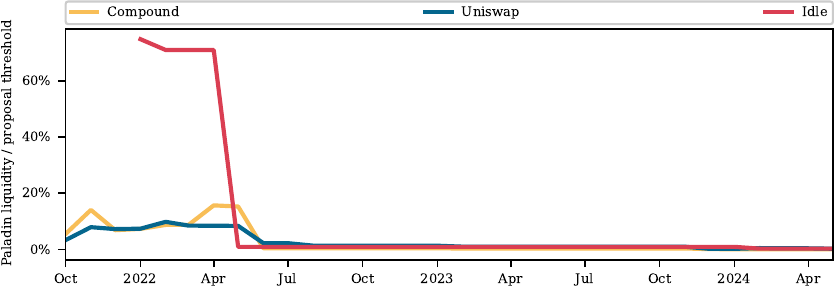}
    \caption{Amount of liquidity available on Paladin in relation to the proposal threshold of the respective protocol.}
    \label{fig:paladin}
\end{figure}

Figure~\ref{fig:paladin} shows the liquidity available on Paladin compared to the proposal threshold for Compound, Uniswap, and Idle over time. Currently, there is nearly no liquidity available on Paladin.

\balance
\clearpage

\end{document}

%% file: attacks.tex
\def\rot{\rotatebox}
\FloatBarrier
\setlength\extrarowheight{-4pt}
\begin{table*}[!h]
\resizebox{1\textwidth}{!}{
\begin{tabular}{>{\scriptsize}r rr c c r *2{C{c} @{}} @{\hspace*{.5cm}} *5{C{c} @{}} @{\hspace*{.5cm}} *5{C{c} @{}} @{\hspace*{.5cm}} *2{C{c} @{}} } & {\scriptsize\rot{90}{date}}& {\scriptsize\rot{90}{blockchain}} & {\scriptsize\rot{90}{attack purpose}}& {\scriptsize\rot{90}{successful}} & {\scriptsize\rot{90}{attack damages}}
                                               & head=bribing holders/delegates (BR1)& head=vote buying protocols (BR2) 
                                               & head=token purchase (TC1)& head=token loan (TC2)  & head=flash loan (TC3) & head=whale activation  (TC4) & head= majority coalition (TC5)
                                               & head=UI issues (HCI1) & head=proposal obfuscation (HCI2)& head=proposal spam (HCI3)  & head=social infiltration (HCI4) &head=behavioral manipualtion (HCI5)
                                               & head=code vulnerability (CP1)& head=protocol vulnerability (CP2)
                                                \\[0.5em]
    
    \multicolumn{1}{l}{\scriptsize\textbf{incidents \& attacks}}  \\
        Audius~\cite{audius2022postmortem,cointelegraph2022audius}&{\scriptsize Jul 2022}&{\scriptsize ETH}& {\scriptsize\faIcon{dollar-sign}} &{\scriptsize\ding{51}}&{\scriptsize\$6.1M}
        & .               & . 
        & .       & .  & . & . & . 
        & .                 & .             & .     & .  & . 
        & technical    & .  
        \\
        Beanstalk~\cite{beanstalk2022}&{\scriptsize Apr 2022}&{\scriptsize ETH}& {\scriptsize\faIcon{dollar-sign}} &{\scriptsize\ding{51}}&{\scriptsize\$182M}
        & .               & . 
        & .       & .  & buy & .  & .
        & .                 & chi             & .     & .  & . 
        & .    & .  
        \\
        BigCap DAO~\cite{2023bigcap} &{\scriptsize Sep 2023}&{\scriptsize ETH}&{\scriptsize\faIcon{dollar-sign}}&{\scriptsize\ding{55}}&
        & .               & .    
         & buy       & .  & . & . & .
        & .                 &       chi       & .     & .  & . 
        & .    & .  
        \\
        Binance~\cite{2022binance}&{\scriptsize Oct 2022}&{\scriptsize ETH}&{\scriptsize\faQuestion}&&
        & .               & . 
          & .       & .   & .      & buy  & .
        & .                 & .             & .     & .  & . 
        & .    & .  
      \\
        Build Finance~\cite{2022buildfinance1,2022buildfinance2,2022buildfinance3,2022buildfinance4,2022buildfinance5} &{\scriptsize Feb 2022}&{\scriptsize ETH}&{\scriptsize\faIcon{dollar-sign}}&{\scriptsize\ding{51}}&{\scriptsize\$470K}
        & .               & . 
          & buy       & .  & . & .  & .
        & .                 &       chi       & .     & .  & . 
        & .    & .  
      \\
        Compound~\cite{2022compound}&{\scriptsize Feb 2022}&{\scriptsize ETH}&{\scriptsize\faIcon{university}} &{\scriptsize\ding{55}}&
        & .               & . 
         & .       & buy    & .    & .  & .
        & .                 & .             & .     & .  & . 
        & .       & .  
          \\
        Curio~\cite{halborn2024curio}&{\scriptsize Mar 2024}&{\scriptsize ETH}&{\scriptsize\faIcon{dollar-sign}} & {\scriptsize\ding{51}}&{\scriptsize \$16M}
        & .               & . 
        & .       & .    & .     & .  & .
        & .                 & .             & .     & .  & . 
        & technical       &  . 
          \\
        Curve A~\cite{2023curvewars} &{\scriptsize ongoing}&{\scriptsize ETH}& {\scriptsize\faIcon{sync-alt}} &&
        & bribe               & . 
         & .       & .     & .    & .   & .
        & .                 & .             & .     & .  & . 
        & .       & technical  
       \\
        Curve B~\cite{2022curveb1,2022curveb2} &{\scriptsize Nov 2021}&{\scriptsize ETH}& {\scriptsize\faIcon{university}} &{\scriptsize\ding{51}}&
        & bribe               & . 
        & .       & .     & .    & .  & .
        & .                 & .             & .     & .  & . 
        & .       & technical 
         \\
        ForceDAO~\cite{halborn2021forcedao}&{\scriptsize Apr 2021}&{\scriptsize ETH}&{\scriptsize\faIcon{dollar-sign}} & {\scriptsize\ding{51}}&{\scriptsize \$367K}
        & .               & . 
        & .       & .    & .     & .  & .
        & .                 & .             & .     & .  & . 
        & technical       &  .      
          \\
        Genesis Alpha~\cite{levi2019genesisalpha} &{\scriptsize Feb 2019}&{\scriptsize ETH}&{\scriptsize\faIcon{dollar-sign}}& {\scriptsize\ding{51}}&{\scriptsize \$90K}
        & .               & . 
         & .       & .    & .     & .  & .
        & .                 & .             & .     & .  & . 
        & technical       &  .     
         \\
        Indexed Finance~\cite{2023indexed}&{\scriptsize Nov 2023}&{\scriptsize ETH}& {\scriptsize\faIcon{dollar-sign}}&{\scriptsize\ding{55}}&
        & .               & .   
         & buy       & .    & .     & .   & .
        & .                 & .             & .     & .  & . 
        & .       & .  
           \\
        Kleros~\cite{kleros2023governor} &{\scriptsize Dec 2023}&{\scriptsize ETH}&{\scriptsize\faIcon{dollar-sign}}&{\scriptsize\ding{55}}&
        & .               & . 
        & .       & .    & .     & .     & .
        & .                 & chi            & .     & .  & . 
        & .       & .  
          \\
        Maker DAO B~\cite{2020makerdaob}&{\scriptsize Oct 2020}&{\scriptsize ETH}& {\scriptsize\faIcon{university}}&{\scriptsize\ding{51}}&
        & .               & .   
        & .       & .     & buy       & .   & .
        & .                 & .             & .     & .   & . 
        & .       & .     
           \\
        Maker DAO C~\cite{2022makerdaoc} &{\scriptsize Jan 2022}&{\scriptsize ETH}&{\scriptsize\faIcon{university}}&{\scriptsize\ding{55}}&
        & .               & . 
        & .& buy   & .    & .  & .
        & .                 & .             & .     & .  & . 
        & .       & .      
           \\
        Mango Markets~\cite{2022maangomarket1,2022maangomarket2,2022maangomarket3}&{\scriptsize Oct 2022}&{\scriptsize SOL} &{\scriptsize\faIcon{dollar-sign}}&{\scriptsize\ding{51}}&{\scriptsize \$47M}
        & .               & .    
        & .& .   & .    & .  & .
        & .                 & .             & .     & .  & . 
        & .       & technical     
           \\
        Paladin Lending~\cite{2023paladin} &{\scriptsize ongoing}&{\scriptsize ETH}& {\scriptsize\faIcon{sync-alt}} &&
        &  .           & bribe 
        & .       & .   & .     & . & .
        & .                 & .             & .     & .  & . 
        & .    & .  
        \\
        Steemit~\cite{2020steemit}&{\scriptsize Feb 2020}&{\scriptsize STEEM} &{\scriptsize\faIcon{university}}&{\scriptsize\ding{51}}&
        & .               & . 
         & buy       & .       & .      & . & .
        & .                 & .             & .     & .   & . 
        & .       & .     
         \\
        Synthetify~\cite{2023synthetify1,2023synthetify2,2023synthetify3}& {\scriptsize Oct 2023}&{\scriptsize SOL}&{\scriptsize\faIcon{dollar-sign}}& {\scriptsize\ding{51}}&{\scriptsize\$230K}
        & .               & . 
        & buy       & .      & .    & . & .
        & .                 & .             & chi     & .  & . 
        & .       & .     
          \\
        Tally~\cite{2023tallybug} &{\scriptsize Apr 2021}&{\scriptsize ETH}&{\scriptsize\faQuestion}&&
        & .               & .   
        & .       & .  & .  & .  & .
        & chi                 & .             & .     & .       & . 
        & .     & .  
         \\
         Temple DAO~\cite{2022temple1,2022temple2,2022temple3}&{\scriptsize Oct 2022} &{\scriptsize ETH}&{\scriptsize\faIcon{dollar-sign}}&{\scriptsize\ding{51}}&{\scriptsize \$2.4M}
        & .               & . 
        & .       & .   & .      & .  & .
        & .                 & .             & .     & .  & . 
        & technical    & .  
         \\
        The DAO~\cite{daian2016thedao, dupont2017thedao, 2023thedao} &{\scriptsize Jun 2016} &{\scriptsize ETH}&{\scriptsize\faIcon{dollar-sign}}&{\scriptsize\ding{51}}&{\scriptsize \$50M}
        & .               & . 
        & .       & .      & .   & .  & .
       & .                 & .             & .     & .  & . 
        & technical     & .  
         \\
          
        Tornado Cash~\cite{behnke_explained_2023}&{\scriptsize May 2023}&{\scriptsize ETH}&{\scriptsize\faIcon{dollar-sign}} &{\scriptsize\ding{51}}&{\scriptsize \$2M}
        & .               & .  
         & buy      & .     & .      & .  & .
        & .                 & chi             & .     & .  & . 
        & .       & .     
         \\
        True Seigniorage Dollar~\cite{2021true1,2021true2} &{\scriptsize Mar 2021}&{\scriptsize BSC}&{\scriptsize\faIcon{dollar-sign}}&{\scriptsize\ding{51}}&{\scriptsize\$16K}
        & .               & .  
        &buy      & .       & .      & .   & .
        & .                 & .             & .     & .   & . 
        & .       & .     
        \\

        Wonderland DAO~\cite{2022wonderland}&{\scriptsize Jan 2022}&{\scriptsize ETH}&{\scriptsize\faIcon{university}}&{\scriptsize\ding{51}}&
        & .               & . 
         & .       & .   & .     & . & .
        & .                 & .             & .     & chi & . 
        & .    & .  
       \\
         Venus~\cite{mandal2021venus}&{\scriptsize Sep 2021}&{\scriptsize BSC}&{\scriptsize\faIcon{university}}&{\scriptsize\ding{55}}& 
         & bribe            & . 
         & .& .   & .     & . & buy
        & .                 & .             & .     & . & . 
        & .    & .  
       \\
      Yam Finance~\cite{decrypt2022yam}&{\scriptsize Jul 2022}&{\scriptsize ETH}&{\scriptsize\faIcon{dollar-sign}}&{\scriptsize\ding{55}}& 
        & .               & . 
         & . & .   & .     & . & .
        & .                 & chi             & .     & . & . 
        & .    & .  
        \\
       Yuan Finance~\cite{yuan2021takeover,etherscan2021yuan}&{\scriptsize Sep 2021}&{\scriptsize ETH}&{\scriptsize\faIcon{dollar-sign}}&{\scriptsize\ding{51}}& {\scriptsize\$282K}
        & .               & . 
         & buy & .   & .     & . & .
        & .                 & .             & .     & . & . 
        & .    & .  
       \\[0.5em]
       
    \multicolumn{1}{l}{\scriptsize\textbf{academic papers \& reports}} \\
        Bandwagon Voting~\cite{yaish2024strategic} &{\scriptsize Feb 2024} &{}& &&
        & .               & . 
         & .& .   & .    & .    & .
        & .                 & .             & .     & .  & chi 
        & .       & .     
        \\
        Dark DAOs~\cite{austgen2023dark,augsten2024dark,daian2018dark} &{\scriptsize Jul 2018}&&&&
        & bribe               & .       
         & .       & .     & .       & .  & .
        & .                 & .             & .     & .   & . 
        & .       & .     
           \\
        Maker DAO A~\cite{gudgeon2020crisis} &{\scriptsize Feb 2020}&{\scriptsize ETH}&&&
        & .               & . 
         & buy       & buy     & buy       & .    & .
        & .                 & .             & .     & .   & . 
        & .       & .     
         \\
        Nexus Mutal~\cite{2020nexusmutal} &{\scriptsize Feb 2020} &{\scriptsize ETH}& &&
        & .               & . 
         & .& .   & .    & .    & .
        & .                 & .             & .     & .  & . 
        & technical       & .   
          \\
        Vote Sniping~\cite{rossello2024blockholders} &{\scriptsize Jan 2024} &{}& &&
        & .               & . 
         & .& .   & .    & .    & .
        & .                 & .             & .     & .  & chi 
        & .       & . 
        \\[0.5em]
    \multicolumn{1}{l}{\scriptsize\textbf{audits}} \\
        Agora~\cite{obront2023agora} &{\scriptsize May 2023}& {\scriptsize OP}&&&
        & .               & .   
        & .       & .     & .       & .    & .
        & .                 & .             & .     & .   & . 
        & technical       & technical     
        \\
        Constitution DAO~\cite{hacken2022constitution} &{\scriptsize Jan 2022}&{\scriptsize ETH} &&&
        & .               & .  
        & .       & .     & .       & .   & .
        & .                 & .             & .     & .   & . 
        & .       & technical     
          \\
        Curve C~\cite{trailofbits2020curve} &{\scriptsize Jul 2020}&{\scriptsize ETH} &&&
        & .               & . 
        & .       & .     & .       & .   & .
        & .                 & .             & chi     & .   & . 
        & technical       &  technical     
          \\
        DAO Maker~\cite{hacken2021daomaker} &{\scriptsize Mar 2021}&{\scriptsize ETH} &&&
        & .               & .  
        & .       & .     & .       & .   & .
        & .                 & .             & .     & .   & . 
        & .       & technical     
          \\
        GameDAO~\cite{certik2021gamedao} &{\scriptsize Aug 2021}& {\scriptsize BSC} &&&
        & .               & . 
        & .       & .     & .       & .     & . 
        & .                 & .             & .     & .   & . 
        & technical       & .     
        \\
        Hoprnet~\cite{chainsecurity2021hopr} &{\scriptsize Jun 2021}& {\scriptsize ETH} &&&
        & .               & . 
        & .       & .     & .       & .      & .
        & .                 & .             & .     & .   & . 
        & technical       & .     
        \\
        Keep3r Network~\cite{statemind2022kp3r} &{\scriptsize Sep 2022}&{\scriptsize ETH} &&&
        & .               & .  
        & .       & .     & .       & .   & .
        & .                 & .             & .     & .   & . 
        & technical       & .     
           \\
        Maker DAO D~\cite{openzeppelin2019makerd} &{\scriptsize May 2019}&{\scriptsize ETH}&&&
        & .               & .  
         & .       & .     & .       & .   & .
        & .                 & .             & .     & .   & . 
        & technical       & .  
           \\
        POA Network~\cite{chainsecurity2018poa} &{\scriptsize Sep 2018}&{\scriptsize ETH}&&&
        & .               & .  
         & .       & .     & .       & .   & .
        & .                 & .             & .     & .   & . 
        & technical       & .  
          \\
        Snapshot X~\cite{chainsecurity2023snapshotx} &{\scriptsize Jul 2023}&{\scriptsize EVM}&&&
        & .               & . 
         & buy       & .     & .       & .    & .
        & .                 & .             & .     & .   & . 
        & technical       & .     
         \\

\end{tabular}}
\caption{Categorization of past attacks and incidents, as well as possible attacks uncovered in academic papers, reports, or audits. For each attack, we indicate its purpose: \faIcon{dollar-sign}~signifies that the purpose of an attack was to extract funds from the DAO, \faIcon{university}~indicates that the goal was a long-term (financial) gain, \faIcon{sync-alt}~denotes an ongoing attack (possibility), and \faQuestion~indicates a (potentially) unintentional incident that exemplified vulnerabilities of DAOs. We further indicate whether the attack was successful where appropriate and if so indicate the financial damage of the attack. Finally, we also highlight which attack vector(s) were used. We proceed similarly for (potential) attacks uncovered in academic papers, reports, or audits. Moreover, we provide a brief summary of each (theorized) attack in Appendix~\ref{app:details}. 
}\label{tab:overview}
\end{table*}
\FloatBarrier

%% file: risks.tex
\def\rot{\Rotatebox{180}}
\renewcommand\theadfont{\normalsize}

\newcommand{\specialcell}[2][c]{%
  \begin{tabular}[#1]{@{}c@{}}#2\end{tabular}}

\begin{sidewaystable}

\settowidth\rotheadsize{\theadfont aligned with data}
 \resizebox{1\textwidth}{!}{

\begin{tabular}{l| rcrrrrrrrrrrrrrr }
\toprule
&  \multicolumn{2}{c}{\textbf{voter apathy (RF1) }}  & \multicolumn{3}{c}{\textbf{gov. token liquidity (RF2)}}  & \multicolumn{2}{c}{\textbf{large treasury (RF3)}} & \multicolumn{3}{c}{\textbf{configuration (RF4)}} &\multicolumn{4}{c}{\textbf{centralization (RF5)}}&\multicolumn{2}{c}{\textbf{code (RF6)}}  \\
 \cmidrule(lr){2-3} \cmidrule(lr){4-6} \cmidrule(lr){7-8} \cmidrule(lr){9-11}  \cmidrule(lr){12-15} \cmidrule(lr){16-17} 
    
\specialcell[b]{\textbf{DAO}}
& \rot{\rothead{avg. votes  in \% of delegated vote  }}& \rot{\rothead{avg. votes  in \% of token supply  }}
& \rot{\rothead{avail. liquidity in \% of proposal thresh.}} &\rot{\rothead{avail. liquidity in \% of delegated votes}} & \rot{\rothead{avail. liquidity in \% of  avg. votes}} 
& \rot{\rothead{treasury value  w/ gov. token in \% of delegated votes}} & \rot{\rothead{treasury value  w/o gov. token in \% of delegated votes}} 
&  \rot{\rothead{proposal  delay in blocks }} &    \rot{\rothead{voting  period  in blocks }} &    \rot{\rothead{timelock delay in blocks}} 
& \rot{\rothead{Nakamoto coefficient of  delegated votes }}& \rot{\rothead{Nakamoto coefficient of  token supply}} & \rot{\rothead{number EOAs  holding more gov. token than  delegated votes}} & \rot{\rothead{guardian}	}
& \rot{\rothead{ownership renounce \cite{certik2024dashboard} }}	& \rot{\rothead{mint function \cite{certik2024dashboard}}}
			 \\

\midrule

Aave &  & 3.21 & 8.74 &  & 4.53 &  &  & 7,200 & 72,000 &  &  & 8 &  & \ding{51} & \ding{55} & \ding{51} \\
Ampleforth & 81.13 & 4.58 & 93.55 & 8.64 & 10.02 & 837.18 & 508.76 & 13,140 & 19,710 & 172,800 & 3 & 5 & 4 & \ding{55} & \ding{55} & \ding{55} \\
ArbitrumCore & & 0.85 & 1,595.42 & & 18.68 &  &  & 21,600 & 100,800 &  & & 6 & & \ding{51} & \ding{51} & \ding{55} \\
ArbitrumTreasury & & 0.72 & 1,595.42 & & 22.19 &  &  & 21,600 & 100,800 &  & & 6 & & \ding{51} & \ding{51} & \ding{55} \\
Babylon & 32.07 & 8.71 & 52.68 & 1.80 & 3.02 & 0.00 & 0.00 & 1 & 45,818 & 86,400 & 4 & 2 & 0 & \ding{55} & \ding{55} & \ding{51} \\
Braintrust & 75.93 & 0.10 & 182,186.84 & 2.92 & 381.12 & 29.83 & 0.00 & 1 & 17,280 & 604,800 & 1 & 15 & 0 & \ding{51} & \ding{55} & \ding{55} \\
Compound & 22.19 & 5.12 & 300.00 & 4.26 & 14.64 & 1.21 & 0.94 & 13,140 & 19,710 & 172,800 & 12 & 12 & 0 & \ding{55} & \ding{55} & \ding{55} \\
Cryptex & 48.02 & 5.79 & 0.01 & 0.00 & 0.00 & 294.88 & 1.06 & 1 & 17,280 & 259,200 & 3 & 3 & 0 & \ding{51} & \ding{55} & \ding{55} \\
Curve &  & 47.46 &  &  & 2.35 &  &  &  & 604,800 &  &  & 2 &  &  & \ding{55} & \ding{55} \\
ENS & 33.91 & 1.47 & 124.80 & 3.01 & 8.49 & 251.78 & 10.21 & 1 & 45,818 & 172,800 & 17 & 1 & 0 & \ding{55} & \ding{55} & \ding{51} \\
Fei & 18.78 & 4.83 & 399.22 & 9.63 & 20.67 & 0.00 & 0.00 & 1 & 13,000 & 86,400 & 14 & 2 & 0 &  & \faQuestion & \faQuestion \\
Gas & 54.85 & 1.08 & 116.82 & 18.61 & 10.78 & 0.00 & 0.00 & 1 & 45,818 & 0 & 2 & 2 & 4 & \ding{55} & \ding{55} & \ding{55} \\
Gitcoin & 34.31 & 2.92 & 217.49 & 4.02 & 11.17 & 416.09 & 40.89 & 13,140 & 40,320 & 172,800 & 3 & 3 & 1 & \ding{55} & \ding{55} & \ding{55} \\
Hifi & 51.80 & 1.98 & 216.31 & 2.11 & 3.87 & 0.00 & 0.00 & 13,140 & 36,000 & 172,800 & 4 & 2 & 3 & \ding{55} & \ding{55} & \ding{55} \\
Hop & 24.07 & 0.42 & 361.06 & 22.42 & 85.17 & 3,396.69 & 0.00 & 1 & 45,818 & 172,800 & 7 & 1 & 0 & \ding{55} & \ding{55} & \ding{51} \\
Idle & 24.82 & 7.88 & 365.90 & 11.61 & 46.46 & 0.00 & 0.00 & 100 & 17,280 & 172,800 & 2 & 6 & 0 & \ding{55} & \ding{55} & \ding{55} \\
InstaDapp & 22.25 & 4.65 & 440.24 & 21.04 & 94.62 & 0.00 & 0.00 & 7,200 & 14,400 & 172,800 & 3 & 3 & 0 &  & \ding{51} & \ding{55} \\
Lido &  & 4.11 &  &  & 8.41 &  &  &  & 21,600 &  &  & 10 &  & \ding{55} & \ding{51} & \ding{55} \\
Maker &  & 15.10 &  &  & 20.94 &  &  &  &  &  &  & 16 &  & \ding{55} & \ding{51} & \ding{51} \\
Optimism & 35.61 & 1.11 & $\infty$ & 3.03 & 5.72 &  &  &  & 259,200 &  & 10 & 3 & 1 & \ding{55} & \ding{55} & \ding{51} \\
Pooh & 1.38 & 0.11 & 4,545.07 & 68.55 & 4,203.75 & 0.00 & 0.00 & 1 & 50,400 & 259,200 & 2 & 35 & 0 & \ding{51} & \faQuestion & \faQuestion \\
Radicle & 37.22 & 3.98 & 95.61 & 9.16 & 24.05 & 489.50 & 25.56 & 1 & 17,280 & 172,800 & 2 & 2 & 0 & \ding{55} & \ding{55} & \ding{55} \\
Silo & 26.26 & 2.09 & 1,183.54 & 0.97 & 5.66 & 161.22 & 0.00 & 128 & 21,000 & 172,800 & 3 & 3 & 0 & \ding{51} & \ding{55} & \ding{51} \\
Strike & 70.15 & 2.00 & 39.50 & 17.00 & 19.65 & 0.00 & 0.00 & 1 & 17,280 & 172,800 & 1 & 2 & 2 & \ding{51} & \ding{55} & \ding{55} \\
Sudoswap & 21.50 & 2.81 & 425.47 & 23.51 & 78.18 & 274.26 & 0.00 & 14,400 & 21,600 &  & 6 & 2 & 0 & \ding{55} & \ding{55} & \ding{51} \\
Uniswap & 25.14 & 4.96 & 325.10 & 1.60 & 6.55 & 186.66 & 0.00 & 13,140 & 40,320 & 172,800 & 16 & 10 & 0 & \ding{55} & \ding{55} & \ding{55} \\
\bottomrule
\end{tabular}}
\vspace{6pt}
\caption{An empirical analysis of the susceptibility of a set of 26 DAOs on the Ethereum blockchain as well as Layer 2s Arbitrum and Optimism to the risk factors presented in Section~\ref{sec:riskfactors}. The data is as of the last block of 31 March 2024 on the respective blockchain and block number delays are indicated according to the underlying chain. Any mention of the average votes refers to an average of the previous five executed votes that were not canceled for each DAO. Additionally, the available liquidity refers to the available liquidity of the respective governance token on Uniswap V2 and Uniswap V3 on Ethereum and Uniswap V3 on Arbitrum and Optimism. Finally, missing entries indicate that the respective risk factor measure does not apply to the DAO, e.g., the risk factor measures related to token delegation are only relevant for DAOs that require delegation.}
\label{tab:risk_factors}
\end{sidewaystable}

%% file: main.bbl
\begin{thebibliography}{100}

\bibitem{venus2021vip42}
{Team Bravo Creation}.
\newblock \url{https://app.venus.io/#/governance/proposal/42?chainId=56}, 2021.

\bibitem{2023bigcapetherscan}
Etherscan bigcap.
\newblock \url{https://etherscan.io/tx/0x7ce953acd59947f8a63f053f4ce3405e53afd4d9d7ef487e755cb4ebd82dba3a}, 2023.
\newblock Ethereum transaction.

\bibitem{2023indexed}
Zack Abrams.
\newblock Indexed dao to distribute remaining treasury after defeating hijack attempts.
\newblock \url{https://www.theblock.co/post/264679/indexed-dao-to-distribute-remaining-treasury-after-defeating-hijack-attempts}, 2023.

\bibitem{2022makerdaoc}
AnnabelTUSD.
\newblock Open letter to the makerdao community from tusd.
\newblock \url{https://forum.makerdao.com/t/open-letter-to-the-makerdao-community-from-tusd/12753/1}, 2022.

\bibitem{araujo2022casting}
Victor Ara{\'u}jo and Malu~AC Gatto.
\newblock Casting ballots when knowing results.
\newblock {\em British Journal of Political Science}, 52(4):1709--1727, 2022.

\bibitem{augsten2024dark}
James Austgen, Andres Fabrega, Sarah Allen, Kushal Babel, Mahimna Kelkar, and Ari Juels.
\newblock Daos must confront dark daos — or fall under their shadow.
\newblock \url{https://initc3org.medium.com/daos-must-confront-dark-daos-or-fall-under-their-shadow-b4c47cb6a1be}, 2024.

\bibitem{austgen2023dark}
James Austgen, Andrés Fábrega, Sarah Allen, Kushal Babel, Mahimna Kelkar, and Ari Juels.
\newblock Dao decentralization: Voting-bloc entropy, bribery, and dark daos, 2023.
\newblock \href {https://arxiv.org/abs/2311.03530} {\path{arXiv:2311.03530}}.

\bibitem{barbereau2023distribution}
Tom Barbereau, Reilly Smethurst, Orestis Papageorgiou, Johannes Sedlmeir, and Gilbert Fridgen.
\newblock Decentralised finance’s timocratic governance: The distribution and exercise of tokenised voting rights.
\newblock {\em Technology in Society}, 73:102251, 2023.
\newblock \href {https://doi.org/10.1016/j.techsoc.2023.102251} {\path{doi:10.1016/j.techsoc.2023.102251}}.

\bibitem{barbereau2022distribution}
Tom~Josua Barbereau, Reilly Smethurst, Orestis Papageorgiou, Alexander Rieger, and Gilbert Fridgen.
\newblock Defi, not so decentralized: The measured distribution of voting rights.
\newblock In {\em Proceedings of the Hawaii International Conference on System Sciences 2022}, page~10, 2022.

\bibitem{2022curveb1}
Rob Behnke.
\newblock Explained: The mochi inu governance hack (november 2021).
\newblock \url{https://www.halborn.com/blog/post/explained-the-mochi-inu-governance-hack-november-2021 }, 2021.

\bibitem{behnke_explained_2023}
Rob Behnke.
\newblock {Explained: The Tornado Cash Hack}.
\newblock \url{https://www.halborn.com/blog/post/explained-the-tornado-cash-hack-may-2023}, May 2023.

\bibitem{behrens2017liquid}
Jan Behrens.
\newblock The origins of liquid democracy.
\newblock {\em The Liquid Democracy Journal on electronic participation, collective moderation, and voting systems}, 5, 5 2017.
\newblock URL: \url{https://liquid-democracy-journal.org/issue/5/The_Liquid_Democracy_Journal-Issue005-02-The_Origins_of_Liquid_Democracy.html}.

\bibitem{bell2021blockchain}
Tom~W Bell.
\newblock Blockchain and authoritarianism: The evolution of decentralized autonomous organizations.
\newblock In {\em Blockchain and Public Law}, pages 90--104. Edward Elgar Publishing, 2021.

\bibitem{2023bigcap}
BIGCAP.
\newblock Community alert! this is scam dao proposal.
\newblock \url{https://twitter.com/BIGCAPProject/status/1697958233204490494}, 2023.
\newblock Twitter post.

\bibitem{beanstalk2022}
Everything Blockchain.
\newblock {Beanstalk Exploit - A Simplified Post-Mortem Analysis}.
\newblock \url{https://medium.com/coinmonks/beanstalk-exploit-a-simplified-post-mortem-analysis-92e6cdb17ace}, 2022.

\bibitem{2022temple2}
BlockSec.
\newblock Twitter post on temple dao attack.
\newblock \url{https://twitter.com/BlockSecTeam/status/1579843881893769222}, 2022.

\bibitem{borge2017pop}
Maria Borge, Eleftherios Kokoris-Kogias, Philipp Jovanovic, Linus Gasser, Nicolas Gailly, and Bryan Ford.
\newblock Proof-of-personhood: Redemocratizing permissionless cryptocurrencies.
\newblock In {\em 2017 IEEE European Symposium on Security and Privacy Workshops (EuroS\&PW)}, pages 23--26. IEEE, 2017.

\bibitem{boringsecurity2023proxy}
{Boring Security}.
\newblock {All About Proxy Contracts}.
\newblock \url{https://boringsecurity.com/articles/all-about-proxy-contracts}, 2023.

\bibitem{buchanan1961majority}
James~M. Buchanan.
\newblock Simple majority voting, game theory, and resource use.
\newblock {\em Canadian Journal of Economics and Political Science}, 27(3):337–348, 1961.
\newblock \href {https://doi.org/10.2307/139591} {\path{doi:10.2307/139591}}.

\bibitem{2022buildfinance2}
BuildFinance.
\newblock Twitter post on governance attack.
\newblock \url{https://twitter.com/finance_build/status/1493223190071554049}, 2022.

\bibitem{buterin2017voting}
Vitalik Buterin.
\newblock Notes on blockchain governance.
\newblock \url{https://vitalik.eth.limo/general/2017/12/17/voting.html}, 2017.

\bibitem{buterin2021voting}
Vitalik Buterin.
\newblock Moving beyond coin voting governance.
\newblock \url{https://vitalik.eth.limo/general/2021/08/16/voting3.html}, 2021.

\bibitem{callander2007bandwagons}
Steven Callander.
\newblock Bandwagons and momentum in sequential voting.
\newblock {\em The Review of Economic Studies}, 74(3):653--684, 2007.

\bibitem{2021true1}
{Certik}.
\newblock Exploiting a smart contract without security vulnerabilities: Analysis of true seigniorage dollar attack event.
\newblock \url{https://www.certik.com/resources/blog/exploitingasmartcontractwithoutsecurityvulnerabilitiesanalysisoftrueseignioragedollarattackevent}, 2021.

\bibitem{certik2021gamedao}
Certik.
\newblock {Security Assessment GameDAO}.
\newblock \url{https://skynet.certik.com/projects/gamedao}, 2021.

\bibitem{certik2023general}
{Certik}.
\newblock {Securing The Web3 World}.
\newblock \url{https://www.certik.com/}, 2023.

\bibitem{certik2024dashboard}
{Certik}.
\newblock {Top DAO Dashboards}.
\newblock \url{https://skynet.certik.com/boards/dao}, 2024.

\bibitem{chainsecurity2018poa}
{ChainSecurity}.
\newblock {Security Audit of POA NETWORK’s Smart Contracts}.
\newblock \url{https://chainsecurity.com/wp-content/uploads/2019/03/ChainSecurity_PoA.pdf}, 2018.

\bibitem{chainsecurity2021hopr}
{ChainSecurity}.
\newblock {Code Assessment of the Hoprnet Token Smart Contracts}.
\newblock \url{https://cdn.prod.website-files.com/65d35b01a4034b72499019e8/6644c996df51a11845ac7de3_210629_HOPR-Token_Smart-Contract-Audit-Report_ChainSecurity_compressed.pdf}, 2021.

\bibitem{chainsecurity2023snapshotx}
{ChainSecurity}.
\newblock {Code Assessment of the Snapshot X Smart Contracts}.
\newblock \url{https://cdn.prod.website-files.com/65d35b01a4034b72499019e8/6645a5f08d64f89be8ee4856_ChainSecurity_PoA_compressed.pdf}, 2023.

\bibitem{2023synthetify3}
coinlive.
\newblock Synthetify suffers \$230,000 loss due to governance failure.
\newblock \url{https://www.coinlive.com/news-flash/298994}, 2023.

\bibitem{cointelegraph2022audius}
{Cointelgraph}.
\newblock {Hacker drains \$1.08M from Audius following passing of malicious proposal}.
\newblock \url{https://cointelegraph.com/news/hacker-drains-1-08m-from-audius-following-passing-of-malicious-proposal}, 2022.

\bibitem{consensys2023bestpractices}
{Consensys}.
\newblock {Ethereum Smart Contract Best Practices}.
\newblock \url{https://consensys.github.io/smart-contract-best-practices/development-recommendations/general/external-calls/}, 2023.

\bibitem{2022bestpractices}
Consensys.
\newblock Ethereum smart contract best practices.
\newblock \url{https://consensys.github.io/smart-contract-best-practices/development-recommendations/}, 2023.

\bibitem{2020steemit}
Tim Copeland.
\newblock Steem vs tron: The rebellion against a cryptocurrency empire.
\newblock \url{https://decrypt.co/38050/steem-steemit-tron-justin-sun-cryptocurrency-war}, 2020.

\bibitem{2022buildfinance4}
Tim Copeland.
\newblock Build finance dao suffers 'hostile governance takeover' loses \$470,000.
\newblock \url{https://www.theblock.co/post/134180/build-finance-dao-suffers-hostile-governance-takeover-loses-470000}, 2022.

\bibitem{daian2016thedao}
Phil Daian.
\newblock Analysis of the dao exploit.
\newblock \url{https://hackingdistributed.com/2016/06/18/analysis-of-the-dao-exploit/}, 2016.

\bibitem{daian2020mev}
Philip Daian, Steven Goldfeder, Tyler Kell, Yunqi Li, Xueyuan Zhao, Iddo Bentov, Lorenz Breidenbach, and Ari Juels.
\newblock Flash boys 2.0: Frontrunning in decentralized exchanges, miner extractable value, and consensus instability.
\newblock In {\em 2020 IEEE Symposium on Security and Privacy (SP)}, pages 910--927, 2020.
\newblock \href {https://doi.org/10.1109/SP40000.2020.00040} {\path{doi:10.1109/SP40000.2020.00040}}.

\bibitem{daian2018dark}
Philip Daian, Tyler Kell, Ian Miers, and Ari Juels.
\newblock On-chain vote buying and the rise of dark daos.
\newblock \url{https://hackingdistributed.com/2018/07/02/on-chain-vote-buying/}, 2018.

\bibitem{2022buildfinance3}
Mike Dalton.
\newblock Build finance dao suffers governance takeover attack.
\newblock \url{https://cryptobriefing.com/build-finance-dao-suffers-governance-takeover-attack/}, 2022.

\bibitem{2020nexusmutal}
Roxana Danila.
\newblock Responsible vulnerability disclosure.
\newblock \url{https://medium.com/nexus-mutual/responsible-vulnerability-disclosure-ece3fe3bcefa}, 2020.

\bibitem{twitter2023yam}
Yam DAO.
\newblock {Twitter post on Yam Finance Redemption}.
\newblock \url{https://twitter.com/YamFinance/status/1619790528752791552}, 2023.

\bibitem{decentraland2024lastminute}
{Decentraland}.
\newblock {Change Gov Mechanism to Mitigate Last-Minute Voting in DAO}.
\newblock \url{https://decentraland.org/governance/proposal/?id=00a79921-2dca-4bde-829e-3a503fc602c2}, 2024.

\bibitem{2023deepdao}
DeepDAO.
\newblock Organizations.
\newblock \url{https://deepdao.io/organizations}, 2023.

\bibitem{dextvl}
Defillama.
\newblock Dexes tvl rankings.
\newblock \url{https://defillama.com/protocols/dexes/Ethereum}, 2023.

\bibitem{2021true2}
True~Seigniorage Dollar.
\newblock {Twitter post on TSD attack}.
\newblock \url{https://twitter.com/TrueSeigniorage/status/1370956726489415683}, 2021.

\bibitem{dotan2023vulnerable}
Maya Dotan, Aviv Yaish, Hsin-Chu Yin, Eytan Tsytkin, and Aviv Zohar.
\newblock The vulnerable nature of decentralized governance in defi.
\newblock In {\em Proceedings of the 2023 Workshop on Decentralized Finance and Security}, DeFi '23, page 25–31, New York, NY, USA, 2023. Association for Computing Machinery.
\newblock \href {https://doi.org/10.1145/3605768.3623539} {\path{doi:10.1145/3605768.3623539}}.

\bibitem{dupont2017thedao}
Quinn DuPont.
\newblock Experiments in algorithmic governance: A history and ethnography of “the dao,” a failed decentralized autonomous organization.
\newblock In {\em Bitcoin and beyond}, pages 157--177. Routledge, 2017.

\bibitem{2022buildfinance5}
Ehterscan.
\newblock {Build Finance}.
\newblock \url{https://etherscan.io/tx/0xf7709b0587d89b9d9b04ca04ce54fdc02a5a30435daf1fb4ba1174486e365c9f }, 2022.
\newblock Ethereum transaction.

\bibitem{2022maangomarket3}
Avraham Eisenberg.
\newblock Twitter post on mango markets.
\newblock \url{https://twitter.com/avi_eisen/status/1581326197241180160}, 2022.

\bibitem{etherscan2021yuan}
{Etherscan}.
\newblock {Yuan Finance}.
\newblock \url{https://etherscan.io/tx/0x4556acce865abe3304eefc7d055112afdcab0d64f838790b46fa0d6dde189c9b}, 2021.
\newblock Ethereum transaction.

\bibitem{eyal2016decentralized}
Ittay Eyal and Emin~Gün Sirer.
\newblock {A Decentralized Escape Hatch for DAOs}.
\newblock \url{https://hackingdistributed.com/2016/07/11/decentralized-escape-hatches-for-smart-contracts/}, 2016.

\bibitem{faife_how_2022}
Corin Faife.
\newblock How to stole an election: {BeanStalk} {DAO} \$80million {FlashLoan} attack study case.
\newblock \url{https://blog.verichains.io/p/how-to-stole-an-election-beanstalk}, 2022.

\bibitem{feichtinger2023shortcomings}
Rainer Feichtinger, Robin Fritsch, Yann Vonlanthen, and Roger Wattenhofer.
\newblock The hidden shortcomings of (d)aos -- an empirical study of on-chain governance.
\newblock In {\em Financial Cryptography and Data Security. FC 2023 International Workshops}, pages 165--185, Cham, 2024. Springer Nature Switzerland.

\bibitem{defiant2023nouns}
Owen Fernau.
\newblock {Nouns NFT Holders Opt To ‘Rage Quit’ Through New Fork}.
\newblock \url{https://thedefiant.io/nouns-nft-holders-opt-to-rage-quit-through-new-forky}, Sep 2023.

\bibitem{ford2002delegative}
Bryan~Alexander Ford.
\newblock Delegative democracy.
\newblock Technical report, EPFL scientific publications, May 2002.
\newblock \url{https://infoscience.epfl.ch/record/265695}.

\bibitem{2020makerdaob1}
William Foxley.
\newblock 'flash loans' have made their way to manipulating protocol elections.
\newblock \url{https://www.coindesk.com/tech/2020/10/29/flash-loans-have-made-their-way-to-manipulating-protocol-elections}, 2020.

\bibitem{fracassi2024decentralized}
Cesare Fracassi, Moazzam Khoja, and Fabian Sch{\"a}r.
\newblock Decentralized crypto governance? transparency and concentration in ethereum decision-making.
\newblock {\em Transparency and Concentration in Ethereum Decision-Making (January 10, 2024)}, 2024.

\bibitem{fritsch2022analyzing}
Robin Fritsch, Marino Müller, and Roger Wattenhofer.
\newblock Analyzing voting power in decentralized governance: Who controls daos?, 2022.
\newblock \href {https://arxiv.org/abs/2204.01176} {\path{arXiv:2204.01176}}.

\bibitem{garimidi2023dao}
Pranav Garimidi, Scott~Duke Kominers, and Tim Roughgarden.
\newblock {DAO governance attacks, and how to avoid them}.
\newblock \url{https://a16zcrypto.com/posts/article/dao-governance-attacks-and-how-to-avoid-them/}, 2023.

\bibitem{glaeser2023cicada}
Noemi Glaeser, Istv{\'a}n~Andr{\'a}s Seres, Michael Zhu, and Joseph Bonneau.
\newblock Cicada: A framework for private non-interactive on-chain auctions and voting.
\newblock {\em Cryptology ePrint Archive}, 2023.

\bibitem{wef2022daos}
David Gogel, Bianca Kremer, Aiden Slavin, and Kevin Werbach.
\newblock Decentralized autonomous organizations: Beyond the hype, 6 2022.
\newblock URL: \url{https://www3.weforum.org/docs/WEF_Decentralized_Autonomous_Organizations_Beyond_the_Hype_2022.pdf}.

\bibitem{wef2023daos}
David Gogel, Bianca Kremer, Aiden Slavin, and Kevin Werbach.
\newblock Decentralized autonomous organization toolkit, 1 2023.
\newblock URL: \url{https://www3.weforum.org/docs/WEF_Decentralized_Autonomous_Organization_Toolkit_2023.pdf}.

\bibitem{gudgeon2020crisis}
Lewis Gudgeon, Daniel Perez, Dominik Harz, Benjamin Livshits, and Arthur Gervais.
\newblock The decentralized financial crisis.
\newblock In {\em 2020 Crypto Valley Conference on Blockchain Technology (CVCBT)}, pages 1--15, 2020.
\newblock \href {https://doi.org/10.1109/CVCBT50464.2020.00005} {\path{doi:10.1109/CVCBT50464.2020.00005}}.

\bibitem{hacken2021daomaker}
{Hacken}.
\newblock {DAO Maker Audit Report}.
\newblock \url{https://hacken.io/audits/dao-maker/}, 2021.

\bibitem{hacken2022constitution}
{Hacken}.
\newblock {Consitution DAO Smart Contract Code Review and Security Analysis}.
\newblock \url{https://wp.hacken.io/wp-content/uploads/2022/01/%D0%A1onstitution-DAO_11012022Audit_Report.pdf}, 2022.

\bibitem{halborn2021forcedao}
{Halborn}.
\newblock {Explained: The ForceDAO Hack (April 2021)}.
\newblock \url{https://www.halborn.com/blog/post/explained-the-forcedao-hack-april-2021}, 2021.

\bibitem{halborn2024curio}
{Halborn}.
\newblock {Explained: The Curio Hack (March 2024)}.
\newblock \url{https://www.halborn.com/blog/post/explained-the-curio-hack-march-2024}, 2024.

\bibitem{2023nouns}
Andrew Hayward.
\newblock {Nouns Fork: Disgruntled NFT Holders Exit With \$27 Million From Treasury}.
\newblock \url{https://decrypt.co/197400/nouns-fork-disgruntled-nft-holders-exit-27-million-from-treasury}, 2023.

\bibitem{Heimbach2023defi}
Lioba Heimbach, Eric Schertenleib, and Roger Wattenhofer.
\newblock {DeFi Lending During The Merge}.
\newblock In {\em {5th Conference on Advances in Financial Technologies (AFT), Princeton, NJ, USA}}, October 2023.

\bibitem{Heimbach2023short}
Lioba Heimbach, Eric Schertenleib, and Roger Wattenhofer.
\newblock {Short Squeeze in DeFi Lending Market: Decentralization in Jeopardy?}
\newblock In {\em {3rd Workshop on Decentralized Finance (DeFi), Bol, Brac, Croatia}}, May 2023.

\bibitem{2022maangomarket1}
Louis Husney.
\newblock Mango markets madness: A case study on the mango markets exploit.
\newblock \url{https://infotrend.com/mango-markets-madness-a-case-study-on-the-mango-markets-exploit/}, 2023.

\bibitem{2023curvewars}
{Jimmy Aki}.
\newblock The curve wars.
\newblock \url{https://www.techopedia.com/definition/the-curve-wars}, 2023.

\bibitem{jordan2003majorityrule}
James~S. Jordan.
\newblock {\em Majority rule with dollar voting}, pages 211--220.
\newblock Springer Berlin Heidelberg, Berlin, Heidelberg, 2003.
\newblock \href {https://doi.org/10.1007/978-3-540-24784-5_13} {\path{doi:10.1007/978-3-540-24784-5_13}}.

\bibitem{kiayias2023sok}
Aggelos Kiayias and Philip Lazos.
\newblock Sok: Blockchain governance.
\newblock In {\em Proceedings of the 4th ACM Conference on Advances in Financial Technologies}, AFT '22, page 61–73, New York, NY, USA, 2023. Association for Computing Machinery.
\newblock \href {https://doi.org/10.1145/3558535.3559794} {\path{doi:10.1145/3558535.3559794}}.

\bibitem{kitzler2023governance}
Stefan Kitzler, Stefano Balietti, Pietro Saggese, Bernhard Haslhofer, and Markus Strohmaier.
\newblock The governance of decentralized autonomous organizations: A study of contributors' influence, networks, and shifts in voting power, 2023.
\newblock \href {https://arxiv.org/abs/2309.14232} {\path{arXiv:2309.14232}}.

\bibitem{kleros2023governor}
{Kleros}.
\newblock {Kleros Blocks Attack on POH Governor, Saves 46 ETH}.
\newblock \url{https://typefully.com/Kleros_io/5yDM4vb}, 2023.

\bibitem{2022temple1}
Oliver Knight.
\newblock Defi protocol temple dao struck by \$2.3m exploit.
\newblock \url{https://www.coindesk.com/business/2022/10/11/defi-protocol-temple-dao-struck-by-23m-exploit/}, 2022.

\bibitem{2023synthetify2}
Jack Kubinec.
\newblock Dao on solana loses \$230k after ‘attack proposal’ goes unnoticed.
\newblock \url{https://blockworks.co/news/solana-exploit-dao-hacker}, 2023.

\bibitem{lan2005sale}
Luh~Luh Lan and Loizos Leracleous.
\newblock Shareholder votes for sale, Jul 2005.
\newblock URL: \url{https://hbr.org/2005/06/shareholder-votes-for-sale}.

\bibitem{2022buildfinance1}
Isabelle Lee.
\newblock A crypto collective lost \$470,000 after one individual amassed enough tokens to take control of the group's treasury.
\newblock \url{https://markets.businessinsider.com/news/currencies/build-finance-dao-treasury-discord-crypto-build-token-metric-2022-2}, 2022.

\bibitem{lee2021gev}
Leland Lee and Ariah Klages-Mundt.
\newblock Governance extractable value.
\newblock \url{https://ournetwork.substack.com/p/our-network-deep-dive-2}, 4 2021.

\bibitem{levi2018arc}
Adam Levi.
\newblock The arc platforms.
\newblock \url{https://medium.com/daostack/the-arc-platform-2353229a32fc}, 2018.

\bibitem{levi2019genesisalpha}
Adam Levi.
\newblock A technical analysis of the genesis alpha hack.
\newblock \url{https://medium.com/daostack/a-technical-analysis-of-the-genesis-alpha-hack-f8e34433c14b}, 2019.

\bibitem{lido2022twophase}
Lido.
\newblock {Moving To Two-Phase Voting}.
\newblock \url{https://blog.lido.fi/moving-to-two-phase-voting/}, 2022.

\bibitem{anonymous2024git}
{Lioba Heimbach}.
\newblock {DAO Vulnerability}.
\newblock \url{https://github.com/liobaheimbach/DAOVulnerability}, 2024.

\bibitem{lloyd2023veToken}
Thomas Lloyd, Daire O'Broin, and Martin Harrigan.
\newblock Emergent outcomes of the vetoken model.
\newblock In {\em 2023 IEEE International Conference on Omni-layer Intelligent Systems (COINS)}, pages 1--6, 2023.
\newblock \href {https://doi.org/10.1109/COINS57856.2023.10189201} {\path{doi:10.1109/COINS57856.2023.10189201}}.

\bibitem{2020makerdaob}
LongForWisdom.
\newblock {[Urgent] Flash Loans and securing the Maker Protocol}.
\newblock \url{https://forum.makerdao.com/t/urgent-flash-loans-and-securing-the-maker-protocol/4901}, 2020.

\bibitem{makerdao2023shutdown}
{Maker}.
\newblock {Maker Protocol Emergency Shutdown}.
\newblock \url{https://docs.makerdao.com/smart-contract-modules/shutdown}, 2023.

\bibitem{2022binance}
Shaurya Malwa.
\newblock Binance denies allegations it intends to use users' uniswap tokens for voting.
\newblock \url{https://www.coindesk.com/tech/2022/10/20/binance-denies-allegations-that-it-intends-to-use-users-uniswap-tokens-for-voting/}, 2022.

\bibitem{mark2016thedao}
Dino Mark, Vlad Zamfir, and Emin~Gün Sirer.
\newblock A call for a temporary moratorium on the dao.
\newblock \url{https://hackingdistributed.com/2016/05/27/dao-call-for-moratorium/}, 2016.

\bibitem{decrypt2021whatissnaphot}
{Matt Hussey}.
\newblock {What is Snapshot? The Decentralized Voting System}.
\newblock \url{https://decrypt.co/resources/what-is-snapshot-the-decentralized-voting-system}, 2021.

\bibitem{meir2020strategic}
Reshef Meir, Kobi Gal, and Maor Tal.
\newblock Strategic voting in the lab: compromise and leader bias behavior.
\newblock {\em Autonomous Agents and Multi-Agent Systems}, 34:1--37, 2020.

\bibitem{messias2024understanding}
Johnnatan Messias, Vabuk Pahari, Balakrishnan Chandrasekaran, Krishna~P. Gummadi, and Patrick Loiseau.
\newblock Understanding blockchain governance: Analyzing decentralized voting to amend defi smart contracts, 2024.
\newblock \href {https://arxiv.org/abs/2305.17655} {\path{arXiv:2305.17655}}.

\bibitem{morton2015exit}
Rebecca~B Morton, Daniel Muller, Lionel Page, and Benno Torgler.
\newblock Exit polls, turnout, and bandwagon voting: Evidence from a natural experiment.
\newblock {\em European Economic Review}, 77:65--81, 2015.

\bibitem{2023mixbytes}
Konstantin Nekrasov.
\newblock {DAO Voting Vulnerabilities}.
\newblock \url{https://mixbytes.io/blog/dao-voting-vulnerabilities#rec506108657}, 2023.

\bibitem{2023synthetify1}
Neodyme.
\newblock Twitter post on synthetify attack.
\newblock \url{https://twitter.com/Neodyme/status/1715149044794655145?s=20}, 2023.

\bibitem{evan2019aragonperil}
Evan~Van Ness.
\newblock {Aragon vote shows the perils of onchain governance}.
\newblock \url{https://evanvanness.com/post/184616403861/aragon-vote-shows-the-perils-of-onchain-governance}, 2019.

\bibitem{obront2023agora}
Zach Obront.
\newblock {Agora Audit Report}.
\newblock \url{https://github.com/voteagora/optimism-gov/blob/main/audits/23-05-12_zachobront.md}, 2023.

\bibitem{openzeppelin2019makerd}
{OpenZeppelin Security}.
\newblock {Technical Description of Critical Vulnerability in MakerDAO Governance}.
\newblock \url{https://blog.openzeppelin.com/makerdao-critical-vulnerability}, 2019.

\bibitem{2022optimismcitizen}
Optimism.
\newblock Citizens’ house overview.
\newblock \url{https://community.optimism.io/docs/governance/citizens-house/}, 2023.

\bibitem{2022optimismtoken}
Optimism.
\newblock Token house history.
\newblock \url{https://community.optimism.io/docs/governance/token-house-history/}, 2023.

\bibitem{2023paladin}
Paladin.
\newblock Documentation.
\newblock \url{https://doc.paladin.vote/}, 2023.

\bibitem{chainlink2022reentrancy}
Zubin Pratap.
\newblock {Reentrancy Attacks and The DAO Hack}.
\newblock \url{https://blog.chain.link/reentrancy-attacks-and-the-dao-hack/}, 2022.

\bibitem{mandal2021venus}
{Rikta Mandal}.
\newblock {Venus Protocol Prevented Hostile Takeover Attempt}.
\newblock \url{https://www.cryptotimes.io/2021/09/18/venus-protocol-prevented-hostile-takeover-attempt/}, 2021.

\bibitem{rossello2024blockholders}
Romain Rossello.
\newblock Blockholders and strategic voting in daos' governance.
\newblock {\em Available at SSRN 4706759}, 2024.

\bibitem{2022maangomarket2}
SEC.
\newblock {SEC Charges Avraham Eisenberg with Manipulating Mango Markets’ “Governance Token” to Steal \$116 Million of Crypto Assets}.
\newblock \url{https://www.sec.gov/news/press-release/2023-13}, 2023.

\bibitem{2023mudus}
Mundus Security.
\newblock {Typical governance vulnerabilities: from DAO building to DAO smart contract audit}.
\newblock \url{https://mundus.dev/blog/typical-dao-and-governance-smart-contracts-vulnerabilities}, 2023.

\bibitem{sharma2023unpacking}
Tanusree Sharma, Yujin Kwon, Kornrapat Pongmala, Henry Wang, Andrew Miller, Dawn Song, and Yang Wang.
\newblock Unpacking how decentralized autonomous organizations (daos) work in practice, 2023.
\newblock \href {https://arxiv.org/abs/2304.09822} {\path{arXiv:2304.09822}}.

\bibitem{2022temple3}
Shashank.
\newblock Temple dao hack analysis.
\newblock \url{https://blog.solidityscan.com/temple-dao-hack-analysis-c96db856322c}, 2022.

\bibitem{2023thedao}
David Siegel.
\newblock {Understanding The DAO Hack}.
\newblock \url{https://www.coindesk.com/learn/understanding-the-dao-attack/}, 2023.

\bibitem{statemind2022kp3r}
{Statemind}.
\newblock {KP3R Vulnerability Report}.
\newblock \url{https://statemind.io/blog/kp3r-vulnerability-report}, 2019.

\bibitem{decrypt2022yam}
{Sujith Somraaj}.
\newblock {Yam Finance Safeguards \$3.1M Treasury From Governance Attack}.
\newblock \url{https://decrypt.co/104848/yam-finance-safeguards-3-1m-treasury-governance-attack}, 2022.

\bibitem{sun2023decentralization}
Xiaotong Sun, Charalampos Stasinakis, and Georigios Sermpinis.
\newblock Decentralization illusion in decentralized finance: Evidence from tokenized voting in makerdao polls, 2023.
\newblock \href {https://arxiv.org/abs/2203.16612} {\path{arXiv:2203.16612}}.

\bibitem{2023tallybug}
Tally.
\newblock {Post mortem and impact summary: Tally voting bug}.
\newblock \url{https://blog.tally.xyz/post-mortem-and-impact-summary-tally-voting-bug-6a12616ce717?gi=3bda9305d9b9}, 2023.

\bibitem{tan2023open}
Joshua~Z. Tan, Tara Merk, Sarah Hubbard, Eliza~R. Oak, Joni Pirovich, Ellie Rennie, Rolf Hoefer, Michael Zargham, Jason Potts, Chris Berg, Reuben Youngblom, Primavera~De Filippi, Seth Frey, Jeff Strnad, Morshed Mannan, Kelsie Nabben, Silke~Noa Elrifai, Jake Hartnell, Benjamin~Mako Hill, Alexia Maddox, Woojin Lim, Tobin South, Ari Juels, and Dan Boneh.
\newblock Open problems in daos, 2023.
\newblock \href {https://arxiv.org/abs/2310.19201} {\path{arXiv:2310.19201}}.

\bibitem{audius2022postmortem}
{Team Audius}.
\newblock {Audius Governance Takeover Post-Mortem 7/23/22}.
\newblock \url{https://blog.audius.co/article/audius-governance-takeover-post-mortem-7-23-22}, 2022.

\bibitem{2022wonderland}
Andrew Thurman.
\newblock How did a former quadriga exec end up running a defi protocol? wonderland founder explains.
\newblock \url{https://www.coindesk.com/tech/2022/01/27/how-did-a-former-quadriga-exec-end-up-running-a-defi-protocol-wonderland-founder-explains/}, 2021.

\bibitem{2022compound}
Andrew Thurman.
\newblock Tron’s justin sun accused of ‘governance attack’ on defi lender compound.
\newblock \url{https://www.coindesk.com/tech/2022/02/04/trons-justin-sun-accused-of-governance-attack-on-defi-lender-compound/}, 2022.

\bibitem{trailofbits2020curve}
{TrailOfBits}.
\newblock {Curve DAO Security Assessment}.
\newblock \url{https://github.com/trailofbits/publications/blob/master/reviews/CurveDAO.pdf}, 2020.

\bibitem{uniswapsourcecode}
{Uniswap}.
\newblock {GovernorBravoDelegate}.
\newblock \url{https://github.com/gettty/uniswap-gov/blob/main/contracts/GovernorBravoDelegate.sol}, 2024.

\bibitem{buterin2014daodefinition}
{Vitalik Buterin}.
\newblock {DAOs, DACs, DAs and More: An Incomplete Terminology Guide}.
\newblock \url{https://blog.ethereum.org/2014/05/06/daos-dacs-das-and-more-an-incomplete-terminology-guide}, 2014.

\bibitem{yaish2024strategic}
Aviv Yaish, Svetlana Abramova, and Rainer B{\"o}hme.
\newblock Strategic vote timing in online elections with public tallies.
\newblock {\em arXiv preprint arXiv:2402.09776}, 2024.

\bibitem{yi2020digix}
Ryan~Youngjoon Yi.
\newblock Digixdao: A divorce story -- a case study for voting systems and cryptonative arbitrage.
\newblock \url{https://blog.coinfund.io/digixdao-divorce-story-6ed74b00e2bd}, 2 2020.

\bibitem{yuan2021takeover}
{Yuan Finance}.
\newblock {Yuan Governance Attack Update and Migration Plan}.
\newblock \url{https://medium.com/yuan-finance/yuan-governance-attack-update-and-migration-plan-3b5d949ab466}, 2021.

\bibitem{2022curveb2}
zefram.eth.
\newblock Twitter post on mochi.
\newblock \url{https://twitter.com/boredGenius/status/1458732732540854276 }, 2021.

\bibitem{zhou2023sok}
Liyi Zhou, Xihan Xiong, Jens Ernstberger, Stefanos Chaliasos, Zhipeng Wang, Ye~Wang, Kaihua Qin, Roger Wattenhofer, Dawn Song, and Arthur Gervais.
\newblock Sok: Decentralized finance (defi) attacks.
\newblock In {\em 2023 IEEE Symposium on Security and Privacy (SP)}, pages 2444--2461. IEEE, 2023.

\bibitem{zou2015doodle}
James Zou, Reshef Meir, and David Parkes.
\newblock Strategic voting behavior in doodle polls.
\newblock In {\em Proceedings of the 18th ACM conference on computer supported cooperative work \& social computing}, pages 464--472, 2015.

\end{thebibliography}
